\documentclass[brazilian,twocolumn,prb,superscriptaddress,amsmath,amssymb,showpacs,floatfix,preprintnumbers]{revtex4-2}
\usepackage[latin9,utf8]{inputenc}
\usepackage{float}
\usepackage{textcomp}
\usepackage{amsmath}
\usepackage{graphicx}
\makeatletter
\usepackage{url}
\usepackage{grffile}
\usepackage{bbold}

\newcommand{\lyxmathsym}[1]{\ifmmode\begingroup\def\b@ld{bold}
  \text{\ifx\math@version\b@ld\bfseries\fi#1}\endgroup\else#1\fi}

\usepackage[printwatermark]{xwatermark}
\usepackage{xcolor}
\usepackage{graphicx}
\usepackage{tikz}

\newsavebox\mybox
\savebox\mybox{\tikz[color=red,opacity=0.15]\node{arXiv Version};}
\newwatermark*[
  allpages,
  angle=45,
  scale=10,
  xpos=-35,
  ypos=40
]{\usebox\mybox}

\@ifundefined{textcolor}{}
{%
 \definecolor{BLACK}{gray}{0}
 \definecolor{WHITE}{gray}{1}
 \definecolor{RED}{rgb}{1,0,0}
 \definecolor{GREEN}{rgb}{0,1,0}
 \definecolor{BLUE}{rgb}{0,0,1}
 \definecolor{CYAN}{cmyk}{1,0,0,0}
 \definecolor{MAGENTA}{cmyk}{0,1,0,0}
 \definecolor{YELLOW}{cmyk}{0,0,1,0}
}

\usepackage{dcolumn}
\usepackage{color}
\DeclareGraphicsExtensions{.png .jpg .pdf}

\usepackage{hyperref}
\hypersetup{
     colorlinks = true,
     linkcolor = blue,
     anchorcolor = blue,
     citecolor = blue,
     filecolor = blue,
     urlcolor = blue
     }
\usepackage{braket}
\usepackage{array}
\usepackage{booktabs}
\usepackage{soul}

\makeatother

\begin{document}
\title{Probing the structure and composition of van der Waals heterostructures using
the nonlocality of Dirac plasmons in the terahertz regime}
\author{I. R. Lavor}
\email{icaro@fisica.ufc.br}
\affiliation{Instituto Federal de Educação, Ciência e Tecnologia do Maranhão, KM-04,
Enseada, 65200-000, Pinheiro, Maranhão, Brazil}
\affiliation{Departamento de F\'{i}sica, Universidade Federal do Ceará, Caixa Postal
6030, Campus do Pici, 60455-900 Fortaleza, Ceará, Brazil}
\affiliation{Department of Physics, University of Antwerp, Groenenborgerlaan 171,
B-2020 Antwerp, Belgium}
\author{L. S. R. Cavalcante}
\affiliation{Department of Chemical Engineering, University of California - Davis, CA, U.S.A}
\author{Andrey Chaves}
\affiliation{Departamento de F\'{i}sica, Universidade Federal do Ceará, Caixa Postal
6030, Campus do Pici, 60455-900 Fortaleza, Ceará, Brazil}
\author{F. M. Peeters}
\affiliation{Department of Physics, University of Antwerp, Groenenborgerlaan 171,
B-2020 Antwerp, Belgium}
\author{B. Van Duppen}
\email{ben.vanduppen@uantwerpen.be}

\affiliation{Department of Physics, University of Antwerp, Groenenborgerlaan 171,
B-2020 Antwerp, Belgium}
\date{\today }
\begin{abstract}
Dirac plasmons in graphene are very sensitive to the dielectric properties of the environment. We show that this can be used to probe the structure and composition of van der Waals heterostructures (vdWh) put underneath a single graphene layer. In order to do so, we assess vdWh composed of hexagonal boron nitride and different types of transition metal dichalcogenides (TMDs). By performing realistic simulations that account for the contribution of each layer of the vdWh separately and including the importance of the substrate phonons, we show that one can achieve single-layer resolution by investigating the nonlocal nature of the Dirac plasmon-polaritons. The composition of the vdWh stack can be inferred from the plasmon-phonon coupling once it is composed by more than two TMD layers. Furthermore, we show that the bulk character of TMD stacks for plasmonic screening properties in the terahertz regime is reached only beyond 100 layers. 
\end{abstract}
\maketitle

\section{Introduction}

Graphene~\citep{Novoselov2004} and other two-dimensional (2D) materials, such as the transition metal dichalcogenides~\citep{Manzeli2017,Novoselov2005} (TMDs), have been intensively investigated due to their unique opto-electronic properties~ \citep{Geim2013,Wang2012,Low2014,Low2016,Ranieri2014,Jariwala2014,Zhang2015,Ju2011,Chen2012,Fiori2014,Mak2016}. The optical response of each material is different due to, e.g., the presence or absence of band gaps~\citep{Yu2015,Mak2010}, the specific type of the electronic structure, and is also influenced by the intrinsic mobility of the electrons themselves~\citep{Neto2009}. The latter is especially important for graphene because it is responsible for the manifestation of so-called plasmons, collective excitations of the 2D electron liquid~\citep{GabrieleGiuliani2008,Maier2007}. It has been shown that graphene plasmons, also called Dirac plasmons, referring to the single-particle energy spectrum of graphene~\citep{Grigorenko2012}, can be supported at mid infra-red \citep{Zhong2015,Schuller2010,Low2014} to terahertz (THz) frequencies~\citep{Polini2016,Low2014,Ju2011,Alonso-Gonzalez2016} and show strong electromagnetic field confinement~\citep{Goncalves2015,Grigorenko2012}. TMDs, on the other hand, due to their large band gap~\citep{Mak2010,Splendiani2010}, behave as dielectrics at low frequencies, thus not supporting plasmons if not extrinsically doped~\cite{Li2015}.

\begin{figure}[!t]
\centering{}\includegraphics[width=1\columnwidth]{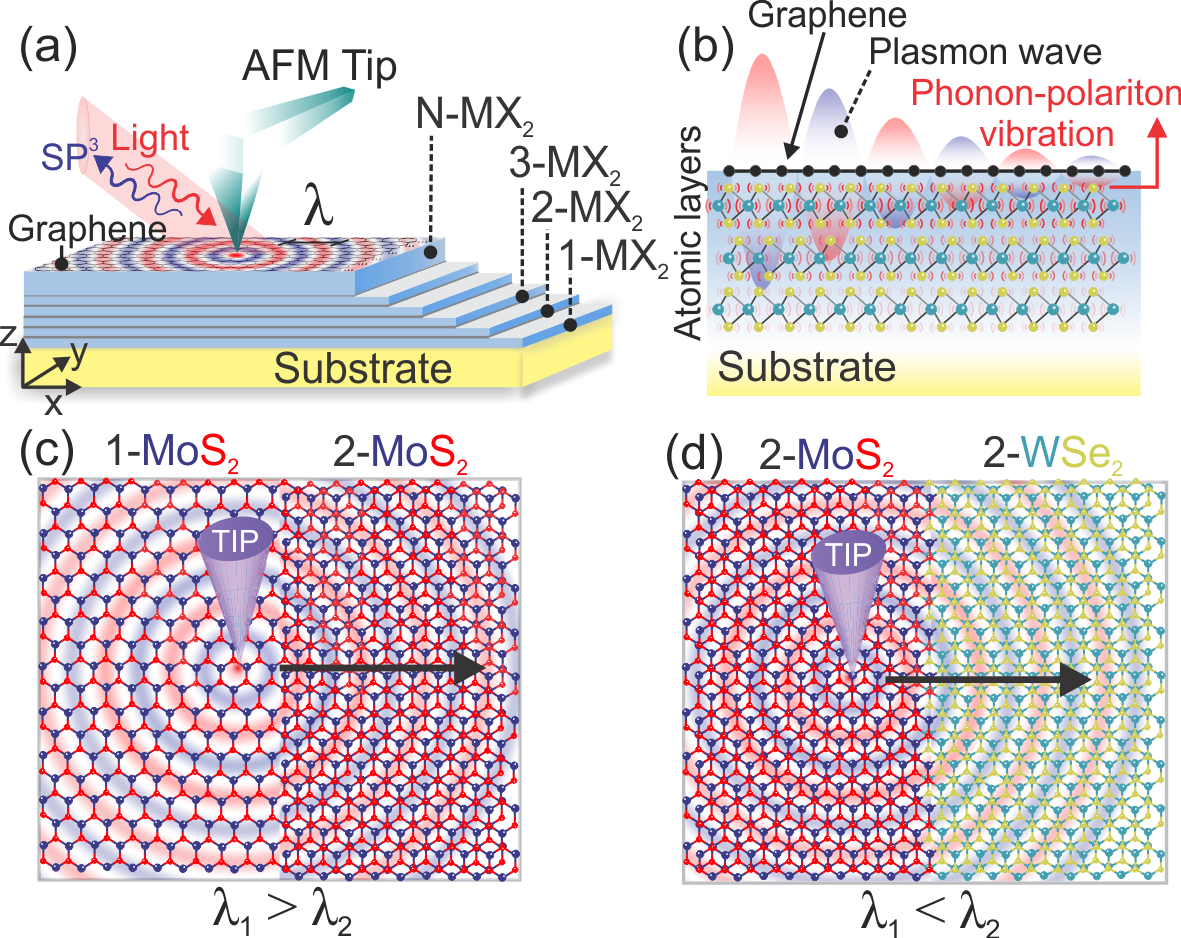}
\caption{(Color online) (a) Schematic illustration of the Dirac plasmon wave in van der Waals heterostructures (vdWh) composed by a monolayer graphene (G) on $\text{N-MX}_{2}$ (M=W,Mo and X=S,Se) and a substrate ($\text{SiO}_{2}$). The graphene surface plasmon-phonon polariton wavelength is $\lambda$. Note that the monolayer graphene covers the entire sample. (b) Illustration of the phonon-polariton vibration in a vdWh composed of $\text{G/3-MX2}_{2}/\text{SiO}_{2}$. Its hybridization with Dirac plasmon originates from the hybridized surface plasmons ($\text{SP}^{3}$). (c) and (d) illustrate the method presented in the paper. $\lambda$ changes when (c) the number of layers in the same material changes, or (d) due to change of materials. $\lambda$ is larger (smaller) when the screening is stronger (weaker). The situation shown in panel (d) occurs for a specific Fermi level and frequency if the phonon frequencies in both materials are different.}
\label{Fig: Ilustration_of_the_GSPP},
\end{figure}

These 2D materials can be combined in so-called van der Waals heterostructures (vdWh)~\citep{Geim2013}. Such structures can be made by stacking different layers on top of each other~\citep{Geim2013,Wang2012,Jariwala2014,Zhang2015,Mak2016,gong2014vertical,Manzeli2017} or even next to each other forming so-called lateral heterostructures~\cite{ozcelik2016band,sahoo2018one,duan2014lateral,gong2014vertical,gong2015two,huang2014lateral,Manzeli2017}. A large corpus of literature has been devoted to the investigation of fabrication techniques to create these nanostructures~\cite{Jariwala2016,ozcelik2016band,sahoo2018one,duan2014lateral,gong2014vertical,gong2015two,huang2014lateral,Geim2013,Wang2012,Novoselov2005,Liu2016,Zhang2015,Manzeli2017}. It has been shown that different opto-electronic properties of the components making up the heterostructure are merged and that by carefully selecting the different constituents, one could achieve materials that are tailor-made to bolster specific behaviour~\citep{Liu2016,Jariwala2016}. Conversely, this means that one could also investigate the opto-electronic response of certain vdWhs to assess their composition and atomic structure. In this paper, we investigate Dirac plasmon states for specific types of vdWh stacks consisting of layers of hexagonal boron nitride (hBN) and different $\text{MX}_{2}$ types of TMDs, composed by a metal (M = Mo or W) layer surrounded by two layers of a chalcogen (X = $\text{S}_{2}$ or $\text{Se}_{2}$), topped by a single graphene layer, as illustrated in Fig. \ref{Fig: Ilustration_of_the_GSPP}(a). Specifically, we investigate the way in which Dirac plasmons in the graphene layer are affected by the constituents of the remainder of the heterostructure and propose a method to infer its local layer number and composition based on local plasmonic properties. Notice that by investigating the effect on Dirac plasmons, we narrow down the spectral region of investigation from the THz to the far-infrared regime. Consequently, other kinds of collective effects, such as excitons, trions or biexcitons, for which traditionally TMDs are well-studied and that have excitation energies of more than 1 eV~\cite{chaves2020bandgap}, will not be affecting the spectral range discussed in this study. Also, we consider doping only in the graphene layer. This means that there are no free electrons in the hBN and TMD layers and, therefore, the plasmonic response can only come from graphene's Dirac plasmons. Consequently, properties such as carrier mobility of the TMD layers are not affecting the obtained results. 

Dirac plasmons in graphene arise as collective excitations of the electronic liquid in the 2D material because of electron-electron Coulomb interaction acting as a restoring force for deformations in the otherwise homogeneous electron density~\citep{Goncalves2015,GabrieleGiuliani2008,Maier2007,Grigorenko2012,Low2014,Low2016}. While the electrons themselves are confined to the 2D material, the electromagnetic field lines associated with the Coulomb force, propagate through the surroundings, and as such are very sensitive to its composition, i.e dielectric properties. Indeed, as shown in Fig.~\ref{Fig: Ilustration_of_the_GSPP}(b), the electromagnetic field is screened due to a polarization of the dielectric environment which effectively slows down plasmon propagation. This combined excitation, i.e. a Dirac plasmon with the surrounding polarization cloud, is often called a surface plasmon-polariton ($\text{SP}^{2}$)~\cite{Griffiths2017}. However, if the plasmon frequency and wavevector match those of intrinsic phonons in the dielectric environment, both modes can hybridize, yielding a combined surface plasmon-phonon-polariton mode  ($\text{SP}^{3}$)~\cite{Grigorenko2012,Luxmoore2014,Wu2016,Dai2015}.

The Dirac plasmon modes, coupled to the underlying heterostructure, can be measured by making use of the well-known scatter-type scanning near-field optical microscope (s-SNOM)~\citep{Lundeberg2017,Lundeberg2016,Alonso-Gonzalez2016} as shown schematically in Fig.~\ref{Fig: Ilustration_of_the_GSPP}(a). This allows to measure the plasmonic wavelength, with a typical resolution of the order of $20$ nm~\citep{Woessner2014,Fei2012,Fei2015,Lundeberg2017,Lundeberg2016,Dai2015,Chen2012,Dai2014}, using interference fringes formed with the plasmon modes scattering of the edge of the heterostructure or at lateral defects in the system. Upon investigation of the dependency of the plasmon wavelength on the tunable graphene carrier concentration, this technique allows to use plasmonic excitations as near-field probes of the material's properties underneath the graphene sheet. In Figs.~\ref{Fig: Ilustration_of_the_GSPP}(c) and (d), we illustrate how this can be used to measure locally the layer structure and composition of the heterostructure.

In this paper, we apply the above-mentioned method to study the dependency of $\text{SP}^{2}$ and $\text{SP}^{3}$ modes on the number and type of layers underneath the graphene sheet. We consider on the one hand hexagonal boron nitride (hBN) and on the other hand four types of TMDs ($\text{WS}_{2}$, $\text{WSe}_{2}$, $\text{MoS}_{2}$ and $\text{MoSe}_{2}$). By using realistic simulations that combine the random phase approximation (RPA) and density functional theory (DFT) calculations, in combination with the quantum electrostatic heterostructure model (QEH)~\citep{Andersen2015}, we are able to investigate the way in which plasmon properties depend on the number of heterostructure layers and the chemical composition of these heterostructures. Furthermore, the use of QEH also allows to properly account for substrate induced effects such as surface phonons that can interfere with the plasmons as well~\cite{Gjerding2020}. We provide a realistic evaluation of the way in which different numbers of layers of the heterostructure screens the electromagnetic field of the plasmon modes and, as such, decreases its wavelength. Also, the QEH allows to assess $\text{SP}^{3}$ modes, which are characteristic of the chemical composition of the TMDs. In this way, one can achieve a layer sensitivity of a single layer and differentiate between different TMDs for heterostructures thicker than 2 layers.

The paper is structured as follows. In Sec.~\ref{sec:THEORETICAL-FRAMEWORK} we introduce the theoretical treatment of Dirac plasmons in vdWhs, introducing substrate effects and the way in which the QEH calculates the role of each layer separately. In Sec.~\ref{sec:RESULTS-AND-DISCUSSIONS} we calibrate the model against known RPA results and experimental results for graphene/hBN heterostructures and discuss the role of the substrate. In Sec.~\ref{sec: probin layer structure and composition} we show how $\text{SP}^{3}$ modes can be used as a means to probe the vdWh layer structure and composition and, finally, in Sec.~\ref{sec:conclusion} we present our conclusions.

\section{Theory of the dielectric response of heterostructures \label{sec:THEORETICAL-FRAMEWORK}}

Dirac plasmons are resonances of the free electron liquid in graphene. These modes can be obtained by solving the plasmon equation which corresponds to the zeroes of the total system's dielectric function $\epsilon(q,\omega)$~\citep{GabrieleGiuliani2008,Fetter2003,Maier2007,Hwang2007,Wunsch2006,Principi2009}
\begin{equation}
    \epsilon(q,\omega) = 1 - v\left(q,\omega\right) \tilde{\chi}_{\rm nn}(q,\omega) = 0~.
    \label{plasmon_equation}
\end{equation}

In Eq.~(\ref{plasmon_equation}), $\tilde{\chi}_{\rm nn}(q,\omega)$ is the proper density-density response function~\cite{GabrieleGiuliani2008} and $v\left(q,\omega\right)$ is the Fourier transform of the Coulomb interaction between the Dirac electrons. In general, both factors depend on the properties of the system as a whole. However, we will approximate the former by the non-interacting density-density response function $\chi^{0}(q,\omega)$, which corresponds to the RPA. This only depends on the properties of graphene. The latter, however, describes electromagnetic field lines that mainly propagate through the surrounding of the graphene sheet, and are, therefore, strongly affected by them. In general, the 2D Fourier transform of the Coulomb interaction is given by
\begin{equation}
   v\left(q,\omega\right)=\frac{2\pi e^{2}}{q\bar{\epsilon}\left(\omega\right)}~.
    \label{coulomb_interaction}
\end{equation}
Equation~(\ref{coulomb_interaction})  makes the role of the heterostructure very clear. Indeed, it is the screening of the Coulomb interaction introduced by the dynamical background dielectric function $\bar{\epsilon}(\omega)$ that encodes the presence of the environment. In order to exemplify how the background dielectrics are affecting the Dirac plasmons, one can calculate the dispersion in the long-wavelength limit and obtain~\citep{Hwang2007,Goncalves2015,Wunsch2006}
\begin{equation}
\lambda(\omega;\bar{\epsilon},E_{\rm F}) = \frac{2\pi}{q(\omega;\bar{\epsilon},E_{\rm F})} = \frac{\pi\alpha_{\rm ee} N_{f} v_{\rm F} }{\hbar\omega^{2}}\frac{E_{\rm F}}{\bar{\epsilon}(\omega)}~.\label{eq:lambda_longwave}
\end{equation}
In Eq.~(\ref{eq:lambda_longwave}), $\alpha_{\rm ee}=2.2$, $N_{f} = 4$ and $v_{\rm F} = 10^{6}$~m/s are parameters related to the graphene sheet corresponding to the graphene fine structure constant, the number of fermion flavours and the Fermi velocity, respectively~\cite{Neto2009}. $E_{\rm F}$ is the Fermi level of graphene. Eq.~(\ref{eq:lambda_longwave}) exemplifies how an increase in the average dielectric constant of the environment decreases the overall plasmon wavelength. As such, since hBN and TMDs all have a larger dielectric screening constant than vacuum, adding more layers to the system should, in general, decrease the wavelength of the collective excitation yielding a screened $\text{SP}^{2}$. However, the environmental dielectric function $\bar{\epsilon}(\omega)$ can have a non-trivial dynamical dependency on $\omega$. This strongly affects the plasmonic wavelength when $\omega$ is close to the frequency of collective lattice vibrations of the environment, such as phonons, which gives rise to the hybrid collective modes $\text{SP}^{3}$. 


In this paper, we consider set-ups as schematically depicted in Figs.~\ref{Fig: Ilustration_of_the_GSPP}(a) and (b), i.e. a system consisting of a substrate, N layers of dielectric such as hBN or $\rm MX_{2}$, and topped with a layer of graphene. We shall denote them as G/N-dielectric/sub. Considering the substrate, we choose to always compare  $\text{SiO}_{2}$. One may also consider other substrates, such as SiC~\cite{liu2010plasmon}, HfO$_2$ and $\text{Al}_{2}\text{O}_{3}$~\cite{ong2012theory}. Our choice for $\text{SiO}_{2}$ as a substrate is motivated as follows: (i) it is widely used in graphene-based plasmon experiments~\cite{Wang2012,Low2016,Jariwala2014,Zhang2015,Ju2011,Chen2012,Alonso-Gonzalez2016,Luxmoore2014,Dai2015,Woessner2014,Fei2012,Fei2015,Dai2014,Fei2011}; (ii) considering  a different substrate, will affect the observed results only in a quantitative way. We do, however, take into account substrate specific effects such as substrate phonons, which will naturally be different for other substrates, but the qualitative result and accuracy of the method will not be affected by this. Both the substrate, as well as the N-layer dielectric, can induce non-trivialities in the environmental dielectric function. In the following, we lay down how to account for both of them. 

\subsection{Coupling to substrate phonons}\label{Coupling plasmons with substrate phonons}

An important non-trivial inclusion of substrate effects are surface phonons. In order to account for them, the most straightforward manner is by considering a frequency-dependent dielectric function of the form~\citep{Luxmoore2014,Goncalves2015}
\begin{equation}
\epsilon_{\rm sub}\left(\omega\right)=\epsilon_{\parallel}^{\infty}+\sum_{n=1}^{M}\frac{f_{n}\omega_{{\rm TO},n}^{2}}{\omega_{{\rm TO},n}^{2}-\omega^{2}-i \omega \gamma_{{\rm TO},n}}~.\label{eq:eps_sub}
\end{equation}
In Eq.~(\ref{eq:eps_sub}), $\epsilon_{\parallel}^{\infty}$ is the in-plane high-frequency dielectric constant, $M$ represent the number of surface transverse optical (TO) phonon modes, and $\omega_{{\rm TO},n}$ and $\gamma_{{\rm TO},n}$ are respectively the frequency and damping of the $n$-th TO surface phonon mode, weighted by $f_{n}$. 
To find the exact plasmon-phonon dispersion, and subsequent the wavelength defined in Eq.~(\ref{eq:lambda_longwave}), it suffices to solve the plasmon equation shown in Eq.~(\ref{plasmon_equation}), where in the absence of a dielectric in-between the substrate and the graphene, $\bar{\epsilon}(\omega) = (\epsilon_{0} + \epsilon_{\rm sub}(\omega))/2$. Note that plasmon, phonon and their hybrid modes also correspond to the maxima of the loss function $L(q,\omega)$, which is defined as
\begin{equation}
    L(q,\omega) = -{\rm Im}\left[\frac{1}{\epsilon(q,\omega)}\right]~.
    \label{equation_lossfunction}
\end{equation}
In the following section, we will include the role of the intermediate dielectric through the use of the QEH model. As it accounts for each layer separately, the output of this model is a loss function. Finally, notice that the $\gamma_{{\rm TO},n}$ coefficients are determined by extrinsic factors, such as impurities~\cite{Principi2013} and defects~\cite{Langer2010} in-between the substrate and the heterostructure. They will result in a spectral broadening of the surface phonons. Since their magnitude depends on the specific set-up~\cite{Luxmoore2014}, in this paper we will not include them\cite{Gjerding2020}.

\subsection{Quantum electrostatic heterostructure model}

The quantum electrostatic heterostructure (QEH)~\citep{Andersen2015} model is used to calculate the non-local dynamical response of the considered vdWh. The model is especially suited for the current investigation because it calculates the dielectric properties of stacks of layers through a bottom-up approach in which the impact of each layer is treated separately.

More recently, the QEH model received an implementation for doped graphene layers in the low energy regime~\cite{Gjerding2020,cavalcante2019}. This regime requires a much more dense grid of $k$-points to correctly describe its properties, which is achieved by the use of an analytical solution for the density response function. The combination of analytical solutions for the response function and DFT calculated induced densities enables more accurate and fast calculations with graphene layers.

The QEH uses the density-density response function of the $i$-th layer $\chi_{i}\left(z,z',\mathbf{q}_{\parallel},\omega\right)$ individually, that was previously obtained through ab-initio calculations. Notice that in this case the vertical spatial dimension $z$ is retained. Subsequently, the total response function of the heterostructure is built by coupling each single layer together by the long-range Coulomb interaction by solving a Dyson-like equation. Omitting the $\mathbf{q}_{\parallel}$ and $\omega$ variables for simplicity, the Dyson equation of the total density-density response function of the complete vdWh reads~\cite{Andersen2015}
\begin{equation}
\chi_{i\alpha,j\beta}=\chi_{i\alpha}\delta_{i\alpha,j\beta}+\chi_{i\alpha}\sum_{k\neq i,\gamma}V_{i\alpha,k\gamma}\chi_{k\gamma,j\beta}~,
\label{Eq. Dyson equation QEH}
\end{equation}
where the Coulomb matrices are defined as
\begin{equation}
V_{i\alpha,k\gamma}\left(\mathbf{q}_{\parallel}\right)=\int\rho_{i\alpha}\left(z,\mathbf{q}_{\parallel}\right)\Phi_{k\gamma}\left(z,\mathbf{q}_{\parallel}\right)dz~,
\end{equation}
and $\Phi_{k\gamma}\left(z,\mathbf{q}_{\parallel}\right)$ is the potential created by the density profile, $\rho_{k\gamma}\left(z,\mathbf{q}_{\parallel}\right)$. In Eq.~(\ref{Eq. Dyson equation QEH}), $\alpha=0,1$ represents the monopole and dipole components, respectively.

Through this formalism, one obtains the inverse dielectric function of the vdWh as
\begin{equation}
\epsilon_{i\alpha,j\beta}^{-1}\left(\mathbf{q}_{\parallel},\omega\right)=\delta_{i\alpha,j\beta}+\sum_{k\gamma}V_{i\alpha,j\beta}\left(\mathbf{q}_{\parallel}\right)\chi_{k\gamma,j\beta}\left(\mathbf{q}_{\parallel},\omega\right)~.
\label{Eq. QEH dielectric function}
\end{equation}
Notice that in contrast to the dielectric function presented in Eq.~(\ref{plasmon_equation}), here we obtain a tensorial form. Consequently, the loss function can be found through 
\begin{equation}
L\left(\mathbf{q}_{\parallel},\omega\right)=-\text{Im}\left[\text{Tr}\left(\epsilon^{-1}\left(\mathbf{q}_{\parallel},\omega\right)\right)\right]~.
\label{TensorialLossfunction}
\end{equation}
Collective modes can now be found as the maxima of this loss function.

Finally, notice that the QEH model also allows to account for intrinsic phonons in the constituent layers. It manages to do so by adding the phonon contribution to the dielectric response function of the individual layers through the calculation of the lattice polarizability, $\alpha_{ij}^{\text{lat}}\left(\omega\right)$, in the optical limit~\cite{Gjerding2020}. This calculation can be considered parameter free, because it is mainly derived from the Born effective charges of the isolated layers~\cite{Resta1994,King-Smith1993}. The Born effective charges are defined as tensors that give the proportionality between the variation of the polarization density due to an atomic displacement and are obtained by the discretized derivative of the 2D polarization density, that follows the Berry phase formalism, in the finite difference method~\cite{Gjerding2020,Resta1994,King-Smith1993}. Thus, considering the contributions of electrons and phonons, the total monopole and dipole component of $i$-th layer are defined as
\begin{subequations}
\begin{equation}
\chi_{i0}^{\text{total}}\left(\mathbf{q}_{\parallel},\omega\right)=\chi_{i0}^{\text{el}}\left(\mathbf{q}_{\parallel},\omega\right)-\mathbf{q}_{\parallel}^{2}\alpha_{\parallel}^{\text{lat}}\left(\omega\right)
\end{equation}
\begin{equation}
\chi_{i1}^{\text{total}}\left(\mathbf{q}_{\parallel},\omega\right)=\chi_{i1}^{\text{el}}\left(\mathbf{q}_{\parallel},\omega\right)-\alpha_{zz}^{\text{lat}}\left(\omega\right)~,
\end{equation}
\end{subequations}
where $\alpha_{\parallel}^{\text{lat}}$ denotes the $2\times2$ in-plane submatrix of $\alpha^{\text{lat}}$. The total response functions are then used in Eq. (\ref{Eq. Dyson equation QEH}), from which the consecutive loss function is obtained. (More details of the QEH model and the way it includes phonons is described in Ref.~\onlinecite{Gjerding2020}).

The major advantage of the use of the QEH model is the availability of a vast database containing the dielectric building blocks of 2D materials~\citep{Link1}, allowing us to reuse previously obtained DFT results. This enables the careful analysis of different vdWh systems on a layer-by-layer basis, without the need to treat the dielectric environment as slabs of bulk material. 

\begin{table}[b]
\centering{}\caption{Phonon parameters of the substrate.
Three optical transverse (TO) phonons were considered for $\text{SiO}_{2}$.
The values of TO frequencies ($\omega_{TO,n}$) and their respective oscillator strength contribution ($f_{n}$) were extracted from Ref. [\onlinecite{Luxmoore2014}].}
\begin{tabular*}{1\columnwidth}{@{\extracolsep{\fill}}>{\centering}m{0.11\textwidth}>{\centering}m{0.09\textwidth}>{\centering}m{0.09\textwidth}>{\centering}m{0.09\textwidth}}
 &  &  & \tabularnewline
\hline 
\hline 
{\centering{}} & $n=1$ & $n=2$ & $n=3$\tabularnewline
\hline 
$\omega_{TO,n}$ (meV) & 55.58 & 98.22 & 139.95\tabularnewline
$f_{n}$ & 0.7514 & 0.1503 & 0.60111\tabularnewline
\hline 
 &  &  & \tabularnewline
\end{tabular*}\label{Table SiO2}
\end{table}

\section{Substrate effects and calibration\label{sec:RESULTS-AND-DISCUSSIONS}}

Before we discuss the impact of the number of layers and composition of the vdWhs on the plasmon properties, we first investigate the role of the substrate on which the total system of vdW coupled layers rest. We assume this substrate to be $\text{SiO}_{2}$, which is frequently used for this purpose \cite{Luxmoore2014, Fei2011, Fei2012, Yan2013}. Furthermore, we use the well-studied hBN-graphene heterostructure as a means to calibrate the QEH code against two sets of experimental results~\cite{Woessner2014,Dai2015}.

\subsection{The importance of substrate surface phonons}

In order to elucidate the effect of the $\text{SiO}_{2}$ substrate and to calibrate the QEH implementation of substrate effects, we consider the environmental dielectric function $\epsilon_{\rm sub}\left(\omega\right)$ as discussed in Sec.~\ref{Coupling plasmons with substrate phonons} both in the RPA treatment and with the QEH model. Table~\ref{Table SiO2} contains the values of the frequency $\omega_{{\rm TO},n}$ and oscillator strength $f_{n}$ of the three TO surface phonons present in $\text{SiO}_{2}$~\citep{Luxmoore2014}. The high-frequency limit of the $\text{SiO}_{2}$ in-plane dielectric constant is $\epsilon_{\parallel}^{\infty}=2.4$. 

\begin{figure}[!t]
\centering{}\includegraphics[width=1\columnwidth]{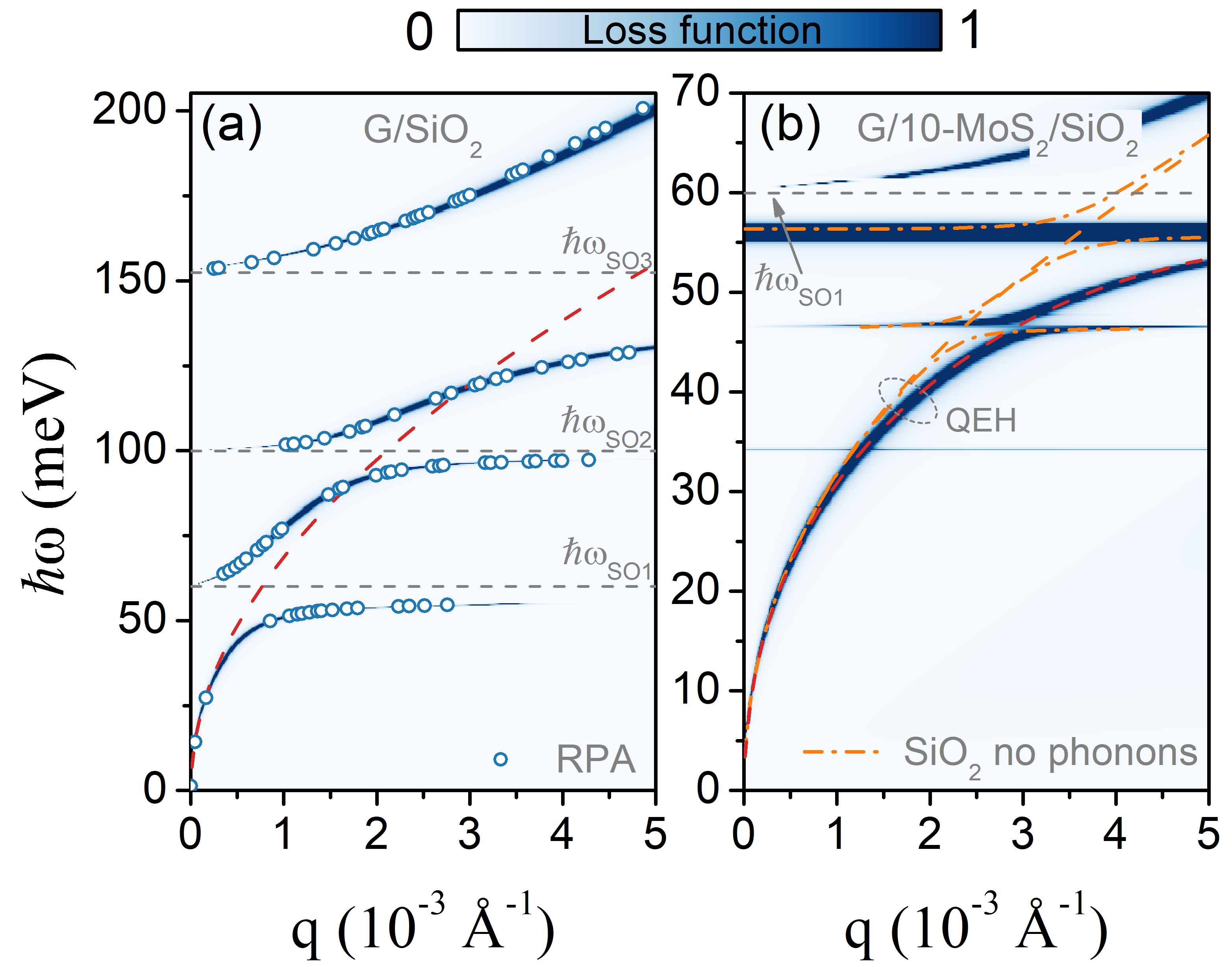}
\caption{(Color online) (a) Comparison between the QEH (loss function) and RPA (symbols) for the $\text{SP}^{3}$ dispersion in graphene with $E_{F}=0.4\text{ eV}$ on $\text{SiO}_{2}$ as a substrate. Hybridization with the substrate phonons ($\hbar \omega_{SOi = 1, 2, 3}$, horizontal gray dashed lines) is clearly visible.  (b) Results at $E_{F}=0.1\text{ eV}$ for $\text{G/10-MoS}_{2}$ on $\text{SiO}_{2}$ with phonons (loss function) and without phonons (dash-dotted orange lines), as calculated the QEH. The unhybridized phonon modes, horizontal blue branches in the loss function, have been omitted for $\text{G/10-MoS}_{2}$ on SiO$_{2}$ without phonons (dash-dotted orange lines). For reference, in (a) and (b), the $\text{SP}^{2}$ dispersion without phonons is presented as dashed lines (red and orange, respectively).}

\label{Fig: Results_role_of_phonons_on_SiO2}
\end{figure}

In Fig.~\ref{Fig: Results_role_of_phonons_on_SiO2}(a) we show that RPA and QEH are in excellent agreement by comparing the loss function of QEH with the exact zeroes of the RPA dielectric function. Here, we assumed a graphene  $E_{F}=0.37\text{ eV}$ on $\text{SiO}_{2}$ with three phonons as indicated by the horizontal lines. There are three regions where the $\text{SP}^{2}$ hybridizes into $\text{SP}^{3}$ modes due to the coupling with the surface phonons of the $\text{SiO}_{2}$ substrate. For reference, we represent in Fig.~\ref{Fig: Results_role_of_phonons_on_SiO2}(a) the $\text{SP}^{2}$ dispersion for $\text{G/SiO}_{2}$ (dashed red curve) with a static dielectric constant $\epsilon_{0}=3.9$~\cite{McPherson2003}.

\begin{figure*}[!t]
\centering{}\includegraphics[width=0.95\textwidth]{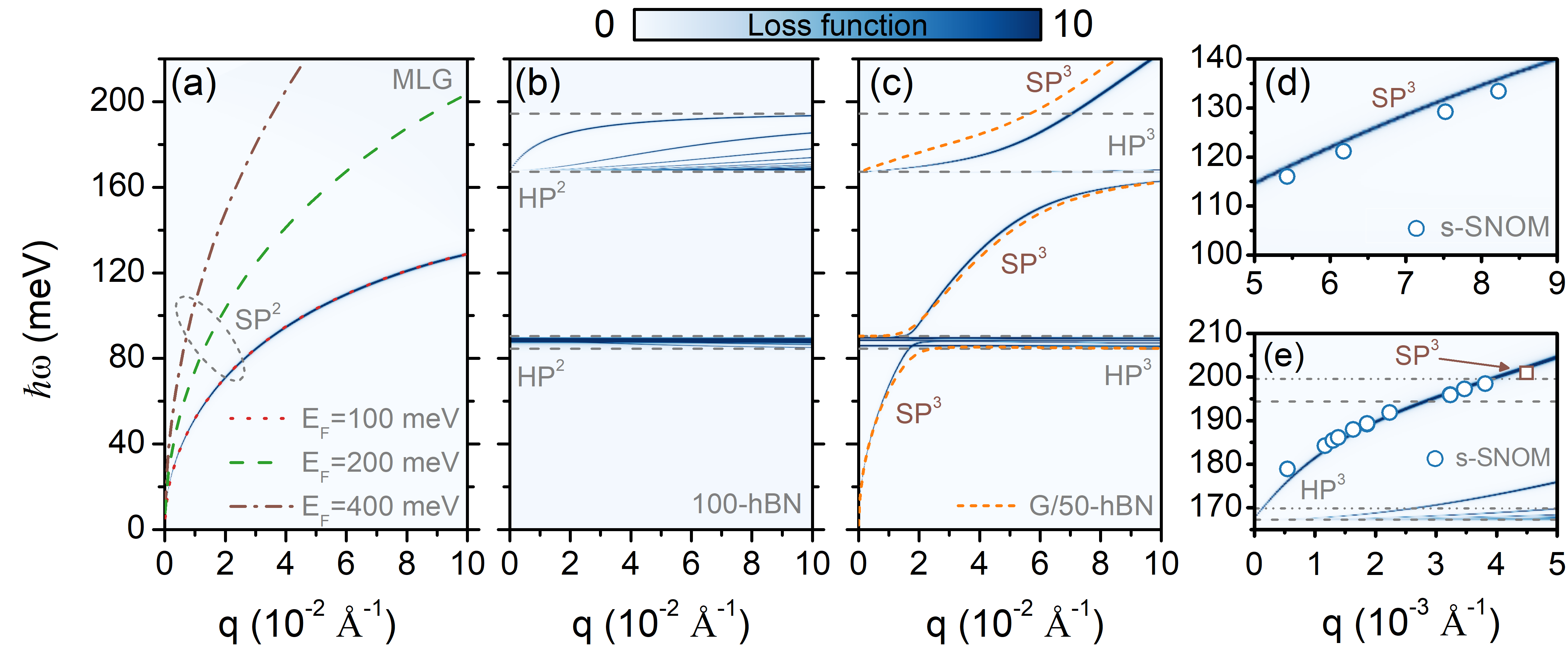}
\caption{(Color online) (a) Plasmon dispersion of the $\text{SP}^{2}$ in free-standing MLG with Fermi energies of $E_{F}=100\text{ meV}$, $200\text{ meV}$ and $400\text{ meV}$. (b) Calculated dispersion of the $\text{HP}^{2}$ in 100 h-BN layers. The hyperbolic regions type I and II correspond to the regions between the two upper and lower grey dashed lines, respectively. (c) Plasmon-phonon dispersion for MLG with $E_{F}=400\text{ meV}$ on 10, represented by the loss function, and 50 h-BN layers with $\text{SiO}_{2}$ (without phonons) as substrate ($\text{G/N-hBN/SiO}_{2}$), represented by the orange dashed lines. (d) and (e) are a comparison between the QEH model and experimental results (symbols)~\citep{Woessner2014,Dai2015} for 21-hBN/G/138-hBN and G/75-hBN, respectively. In panel (e), the RS band II, obtained from the QEH, is delimited by the dashed gray lines (for comparison, the horizontal gray dashed dotted-dotted lines obtained from Ref [\onlinecite{Cai2007}] using solely first principles calculations is used as reference in (e)). The experimental data used in (d) and (e) were extracted from Refs. [\onlinecite{Woessner2014}] and [\onlinecite{Dai2015}]. A false color map represents the loss function in arbitrary units.}
\label{Fig: Results_SP2_hBN_experimental}
\end{figure*}

The inclusion of substrate phonons is important in studying plasmon properties of vdWhs. This is shown in Fig.~\ref{Fig: Results_role_of_phonons_on_SiO2}(b), where we used the QEH model for a vdWh with ten layers of $\text{MoS}_{2}$ with and without substrate phonons. 
This reflects directly on the $\text{SP}^{3}$ dispersion, where in addition to coupling with two of the $\text{MoS}_{2}$ phonons, the plasmons will also couple with the $\text{SiO}_{2}$ surface phonons. Notice that the influence on the $\text{SP}^{3}$ mode is not only manifested at the first phonon frequency $\hbar \omega_{\rm SO1}$, but results in an up to $27\%$ decrease in plasmon wavelength at $\hbar\omega=50\text{ meV}$, i.e. increase in plasmon wave vector $q$ even at much lower frequencies.

\subsection{Calibration of the QEH model for G/N-hBN vdWhs\label{subsec:Plasmon-phonons-in-hBN}}

Using the QEH, we shown in Fig. \ref{Fig: Results_SP2_hBN_experimental}(a) the dispersion of $\text{SP}^{2}$ modes for different values of the Fermi energy $E_{F}$. In the non-retarded regime, these parabolic curves are described by the simple equation~\citep{Hwang2007,Goncalves2015,Wunsch2006} $\omega\sim\sqrt{E_{\rm F} q/\epsilon}$. In Fig.~\ref{Fig: Results_SP2_hBN_experimental}(b), we show the QEH loss function in the absence of a doped graphene sheet for a system containing 100 hBN layers, i.e. a slab of about 33.3 nm thick~\cite{Wagemann2020}. We see the presence of so-called hyperbolic phonon polaritons ($\text{HP}^{2}$) that appear in two given energy bands due to the anisotropy of the hBN dielectric tensor~\citep{Zhang2015,Dai2014}. The two hyperbolic regions, denominated as Reststrahlen (RS) bands, are defined as energy regions where one of the coefficients of the dielectric tensor becomes negative. Due to the fact that these modes are trapped inside the hBN slab, discretization of energy appears.

Upon the addition of a doped graphene sheet, the $\text{SP}^{2}$ modes can hybridize with the $\text{HP}^{2}$ modes of the hBN material, giving rise to new mixed $\text{SP}^{3}$ and hyperbolic plasmon-polariton ($\text{HP}^{3}$) modes as presented in Fig. \ref{Fig: Results_SP2_hBN_experimental}(c) for G/10-hBN and G/50-hBN with $\text{E}_{\text{F}}=400\text{ meV}$. We point out that due to the hyperbolicity of the $\text{HP}^{2}$ modes, the wavelength dependence on the number of hBN layers is opposite for the upper branch of the $\text{SP}^{3}$ modes with respect to the lower branch. This comes as a surprise, since one would expect that screening for a thicker hBN slab should be more important than for a thinner one. However, this observation underlines the difference of $\text{HP}^{2}$ modes with respect to normal phonon polariton modes as presented in the $\rm MX_{2}$ examples in the next section. Also, notice that upon comparison of the hBN results with the $\text{MoS}_{2}$ results presented in Fig.~\ref{Fig: Results_role_of_phonons_on_SiO2}(b), one can see that in the absence of hyperbolicity, no confined modes appear.

A comparison of the results obtained from the QEH and those obtained experimentally for the $\text{SP}^{3}$ and $\text{HP}^{3}\text{(II)}$ modes (the experimental data were extracted from Refs. [\onlinecite{Woessner2014}] and [\onlinecite{Dai2015}], respectively) is illustrated in Figs.~ \ref{Fig: Results_SP2_hBN_experimental}(d)-(e) and shows very good agreement. Figure~\ref{Fig: Results_SP2_hBN_experimental}(d) shows the $\text{SP}^{3}$ dispersion for graphene encapsulated by hBN (21-hBN/G/138-hBN) and Fig.~\ref{Fig: Results_SP2_hBN_experimental}(e) presents the results for G/75-hBN. Notice that upon comparison to the literature, it becomes clear that the exact spectral position of the RS bands is not yet uniquely determined. In the Appendix, we compare the QEH model to different definitions and show our obtained results for the frequencies that define the two RS in hBN, as well as the phonon frequencies for a free-standing monolayer of all TMDs considered in this paper.

\section{Probing layer structure and composition}\label{sec: probin layer structure and composition}

Now, we are in a position to show how one can use $\text{SP}^{2}$ and $\text{SP}^{3}$ modes to probe the layer structure and the composition of the vdWhs. To do so, we assess the plasmon-phonon dispersion of four types of TMDs, namely $\text{MoS}_{2}$, $\text{MoSe}_{2}$, $\text{WS}_{2}$ and $\text{WSe}_{2}$. These materials are often used in the construction of vdWhs~\cite{Liu2016,Jariwala2016,Geim2013,Dai2014,Wang2012,Jariwala2014,Zhang2015a,Mak2016}. Moreover, their chemical similarity makes them interesting candidates to show the sensitivity of the proposed approach. Finally, because of their shared crystallographic structure, namely they all have a $\rm MX_{2}$ form, the number of phonon modes in the 2D materials is all the same, but their respective phonon frequencies differ. For reference, the phonon frequencies of freestanding monolayer for each TMDs used in this paper, obtained from the QEH, are provided in Table \ref{Tab: MX2-AppendixA} of the Appendix.

\subsection{Probing the number of layers}

\begin{figure}[!t]
\centering{}\includegraphics[width=0.9\columnwidth]{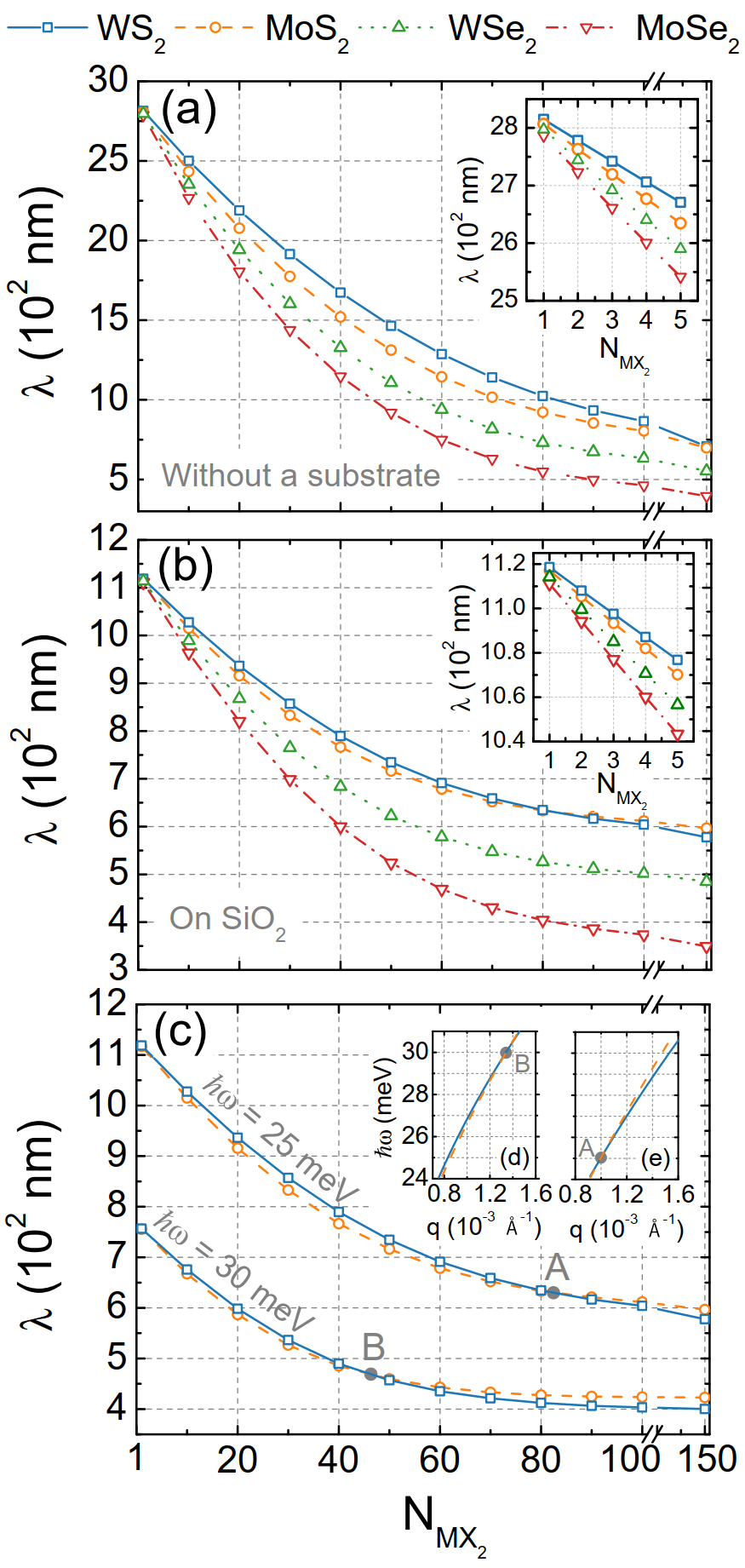}
\caption{(Color online) Graphene surface plasmon-polariton wavelength as a function of the number of layers of $\text{MoSe}_{2}$ (red), $\text{WSe}_{2}$ (green), $\text{MoS}_{2}$ (orange) and $\text{WS}_{2}$ (blue) for $\hbar\omega=25~\text{meV}$ ($\approx$\hspace{0.5mm}6~THz) at $\text{E}_{\text{F}}=100\text{ meV}$, (a) without a substrate and (b) with $\text{SiO}_{2}$ as a substrate. The insets in (a) and (b) show a magnification of the the results for 1 to 5 layers of $\text{MX}_{2}$ without and with a substrate, respectively. (c) Comparison between the $\text{SP}^{3}$ wavelength for $\text{G/N-MoS}_{2}\text{/SiO}_{2}$ and $\text{G/N-WS}_{2}\text{/SiO}_{2}$ at frequencies 25 and 30 meV, respectively. The crossing between the plasmon wavelengths at frequency 30 meV (25 meV) is represented by a grey circle labeled B (A). Inset (d) shows a comparison between the plasmon dispersions for G/44-$\text{MoS}_{2}$/$\text{SiO}_{2}$ and G/44-$\text{WS}_{2}$/$\text{SiO}_{2}$. Inset (e) is the same as (d) but now considering 82 layers of $\text{WS}_{2}$ and $\text{MoS}_{2}$.}
\label{Fig: Results_MLG_on_MX2_with_Phonons_ALL}
\end{figure}

In Fig.~\ref{Fig: Results_MLG_on_MX2_with_Phonons_ALL} we show how the addition of individual $\rm MX_{2}$ layers affects the Dirac plasmon wavelength $\lambda$ for each structure. We fixed the graphene doping at $E_{\rm F} = 100~{\rm meV}$ and excitation frequency $\hbar \omega = 25~{\rm meV}$. The latter is chosen to be below all phonon frequencies in both the substrate and the different $\rm MX_{2}$ layers. In this way, we mainly excite $\text{SP}^{2}$ modes and the effect should be mainly attributed to an increase in dielectric screening due to the permittivity of the $\rm MX_{2}$ layers. 

Panel (a) in Fig.~\ref{Fig: Results_MLG_on_MX2_with_Phonons_ALL} shows how the plasmon wavelength decreases with the number of $\rm MX_{2}$ layers added when no substrate is considered. Also, we show the 'bulk' limit, which is achieved only at about 150 layers. Results for $N > 150$ are verified to be the same (within numerical accuracy) up to 350 $\rm MX_{2}$ layers, thus confirming this bulk limit. This is a surprisingly large number of layers. It was previously established that multilayered structures, such as graphite~\cite{Partoens2006}, achieve their bulk electronic properties at about 10 layers. However, here we show that this does not work for the plasmonic properties, where at least hundreds of layers are needed for bulk behaviour to occur. This observation underlines the necessity for a realistic modeling of plasmon properties, as performed with the QEH model. In Fig.~\ref{Fig: Results_MLG_on_MX2_with_Phonons_ALL}(b), we show how a $\text{SiO}_{2}$ substrate affects the layer dependency. As expected, the substrate results in an overall screening and the wavelength is reduced. Also here, bulk TMD behaviour is reached for about 150 layers. 

Notice that, interestingly, the order of the wavelength values of $\text{WS}_{2}$ and $\text{MoS}_{2}$ is switched when increasing the number of layers. Indeed, while for a few layers, the wavelength in the $\text{WS}_{2}$ system is the largest, in the bulk case, it is the $\text{MoS}_{2}$ system that has the largest wavelength. To describe this peculiar effect, in Fig.~\ref{Fig: Results_MLG_on_MX2_with_Phonons_ALL}(c), we show both results for two different frequencies. As one can see, for a frequency closer to the first phonon frequency of $\text{MoS}_{2}$ (see Tab.~\ref{Tab: MX2-AppendixA}), for example $\hbar\omega=30\text{ meV}$, the crossing occurs for a smaller number of layers than for $\hbar\omega=25\text{ meV}$. This is a direct consequence of the presence of phonons in $\text{MoS}_{2}$. The lowest of them has a frequency given by 34 meV. In $\text{WS}_{2}$, the phonons at 36 meV are not significantly hybridized into $\text{SP}^{3}$ modes, causing this crossing between the plasmon wavelength of these two TMDs. A comparison between the plasmon dispersions for 44 and 82 layers of $\text{WS}_{2}$ and $\text{MoS}_{2}$ is depicted in Figs.~\ref{Fig: Results_MLG_on_MX2_with_Phonons_ALL}(d) and \ref{Fig: Results_MLG_on_MX2_with_Phonons_ALL}(e), respectively, where one can see the crossings at 30 meV (d) and 25 meV (e).

\begin{figure}[!t]
\centering{}\includegraphics[width=1\columnwidth]{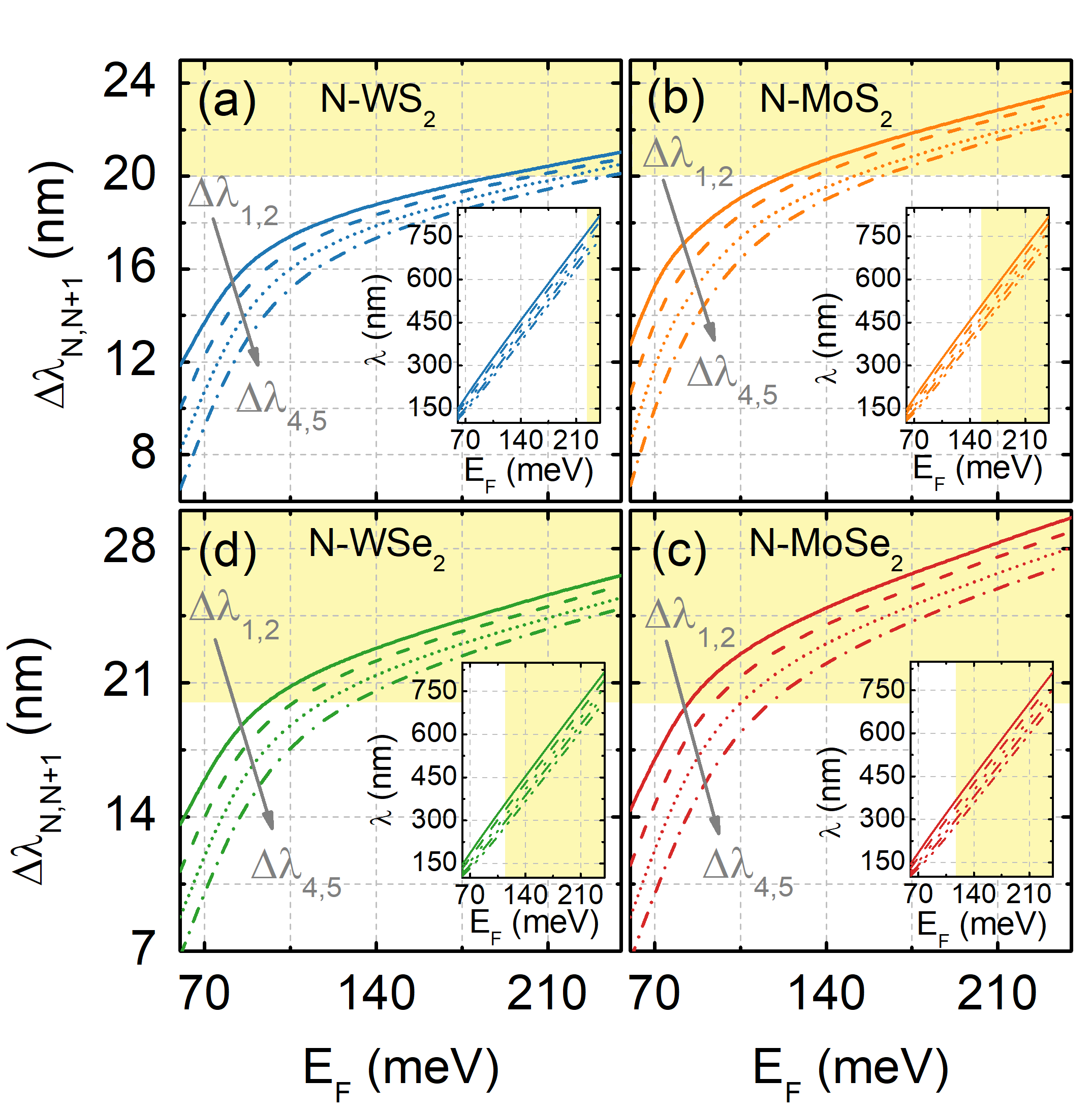}
\caption{(Color online) Difference in the wavelength ($\Delta\lambda_{N,N+1}$) at $\hbar\omega=65\text{ meV}$ ($\approx$\hspace{0.5mm}15.7 THz) between numbers of layers $N$ and $N+1$, from $N=1$ to 5, for (a) $\text{G/N-WS}_{2}\text{/SiO}_{2}$, (b) $\text{G/N-MoS}_{2}\text{/SiO}_{2}$, (c) $\text{G/N-MoSe}_{2}\text{/SiO}_{2}$ and (d) $\text{G/N-WSe}_{2}\text{/SiO}_{2}$. Inset is the wavelength as a function of Fermi level $E_{F}$ at the same frequency. Yellow regions corresponds to $\Delta\lambda_{N,N+1}\protect\geq20\text{ nm}$.}
\label{Fig: Results_Lambda_vs_Ef}
\end{figure}

Finally, in Fig.~\ref{Fig: Results_Lambda_vs_Ef} we show the sensitivity of the proposed method with respect to the number of MX$_2$ layers. As a function of the induced Fermi level in the graphene layer, we show the difference in plasmon wavelength between structures that differ only by one layer. Assuming a lower threshold of $20~{\rm nm}$ for the wavelength resolution, we see that for Fermi levels of more than 140 meV, we can achieve single-layer resolution for every considered TMD-based vdWh.

\subsection{Probing vdWh composition}

\begin{figure*}
\centering{}\includegraphics[width=1\textwidth]{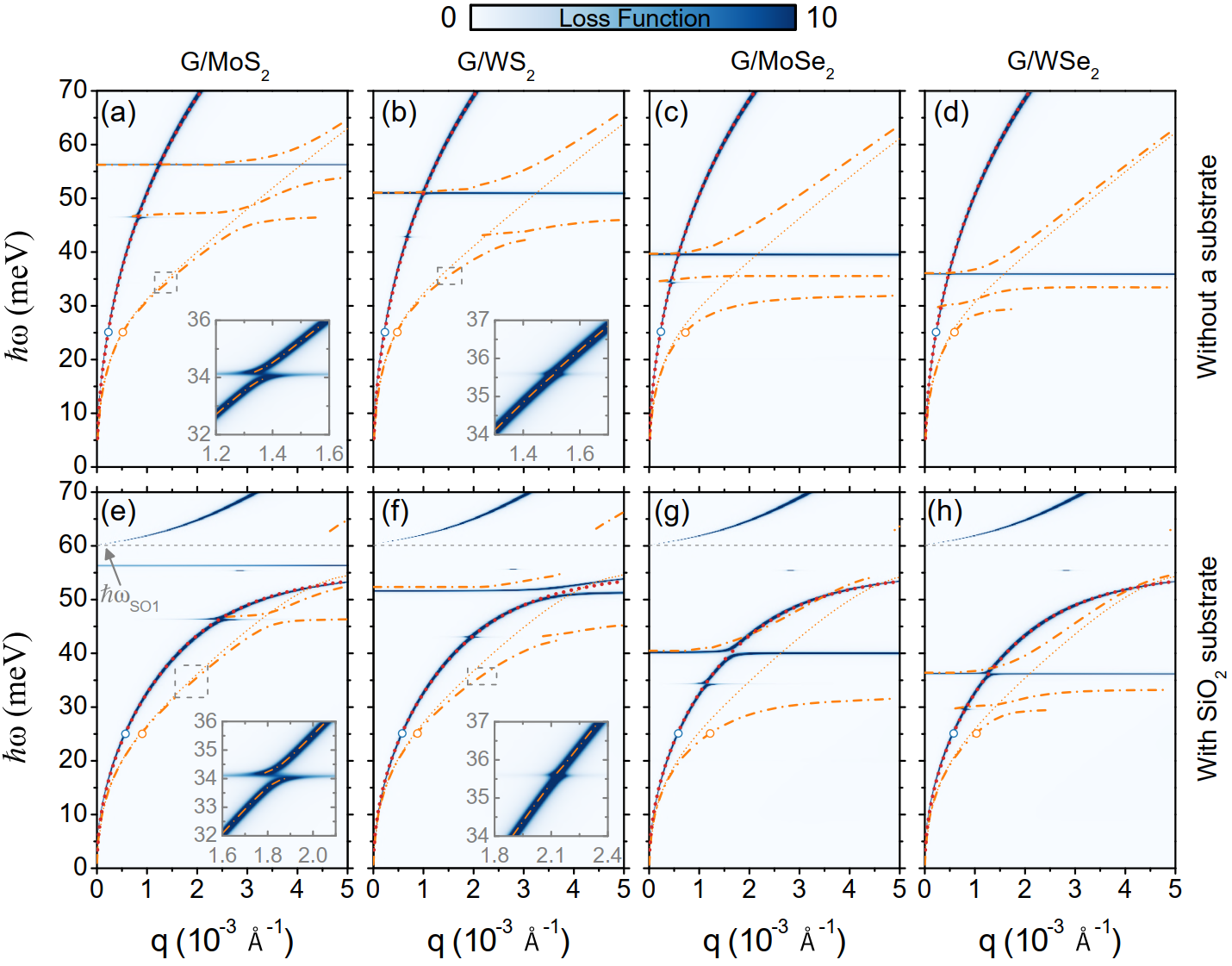}
\caption{(Color online) Overview of the frequency and wave vector dependence of the plasmon $\text{SP}^{3}$, at $E_{F}=100$~meV, for vdWhs with 1 and 50 TMD layers. The background shading is the loss function for $N=1$ TMD layers. The orange curves correspond to $N=50$. The different TMDs under consideration are indicated on the top of each column. The top row are for vdWhs without a substrate, while for the bottom row they are positioned on top of a SiO$_{2}$ substrate. The inset in (a), (b), (e) and (f) show magnifications around the anti-crossing. The red and orange dotted curves denote the SP$^{2}$ modes, for reference.}
\label{Fig: Dispersion_MLG_MX2_ALL}
\end{figure*}

In order to assess the difference between several $\rm MX_{2}$ structures, in Fig.~\ref{Fig: Dispersion_MLG_MX2_ALL} we show the full loss function for the four considered vdWHs, accounting for substrate and TMD phonons. The number of layers remains constant $N_{\text{MX}_{2}}=1$ (loss function) and 50 (orange lines), and we consider free-standing structures and those on a SiO$_2$ substrate. 

The loss function presented in the $(q,\omega)$-plane shows more insight in the behaviour of the different $\text{SP}^{3}$ modes than only through calculating the experimentally relevant wavelength. Indeed, in the different panels of Fig.~\ref{Fig: Dispersion_MLG_MX2_ALL} one can not only distinguish the way in which an increase in the number of vdWhs layers increases the wavevector $q$ (hence decreasing the wavelength $\lambda$), but also verify that each TMD structure bears its own spectrum of phonons. These phonons are the ones that hybridize with the Dirac plasmons and form the $\text{SP}^{3}$ modes and, by investigating the specific type of hybridization, one can infer the chemical properties of the vdWh under consideration. In general, the MX$_{2}$ type TMDs considered in this paper have, in their monolayer form, three acoustic and six optical phonon modes~\cite{Zhang2015a,Zhao2013,MolinaSanchez2011,Peng2016,Berkdemir2013,Sengupta2015}. However, because of the long-wavelength character of the discussed modes, we only excite optical ones at the frequencies considered here. Furthermore, due to symmetry considerations, two pairs of modes are degenerate in the $q\rightarrow 0 $ limit. More details about the phonon structure of these materials are laid down in Appendix~\ref{sec:Appendix:-phonons-frequencies}.

First, we scrutinize the top row of Fig.~\ref{Fig: Dispersion_MLG_MX2_ALL}, in which there is no substrate. Although such TMDs have four distinct optical phonon frequencies, as presented in Tab.~\ref{Tab: MX2-AppendixA} of the Appendix, not all modes are strongly coupled to plasmons.
Typically, the highest energy mode, i.e. the $A_{2}''$ mode, is strongly active in the loss function. However, it is clear that the Dirac plasmon mode also interferes with the other three phonon modes. Furthermore, this interference becomes much more pronounced as the thickness of the TMD stack is increased. This is shown in the insets of panels (a) and (b), where small hybridization with the $E''$ modes is shown.

In Fig.~\ref{Fig: Dispersion_MLG_MX2_ALL}(b) we show the results for a $\rm WS_{2}$ heterostructure. While the A$_{2}''$ mode is more pronounced, the plasmon-phonon hybridization of the other three modes is significantly smaller. Notice that the spectral width of the top mode is also broader than in the case of the $\rm MoS_{2}$ stack. In Figs.~\ref{Fig: Dispersion_MLG_MX2_ALL}(c) and \ref{Fig: Dispersion_MLG_MX2_ALL}(d), the chemical composition of the TMD stacks is changed with a replacement of the sulfur atoms by selenium. Again, a typical phonon spectrum is present, yielding specific types of $\text{SP}^{3}$ modes.

\begin{figure}[!t]
\centering{}\includegraphics[width=0.9\columnwidth]{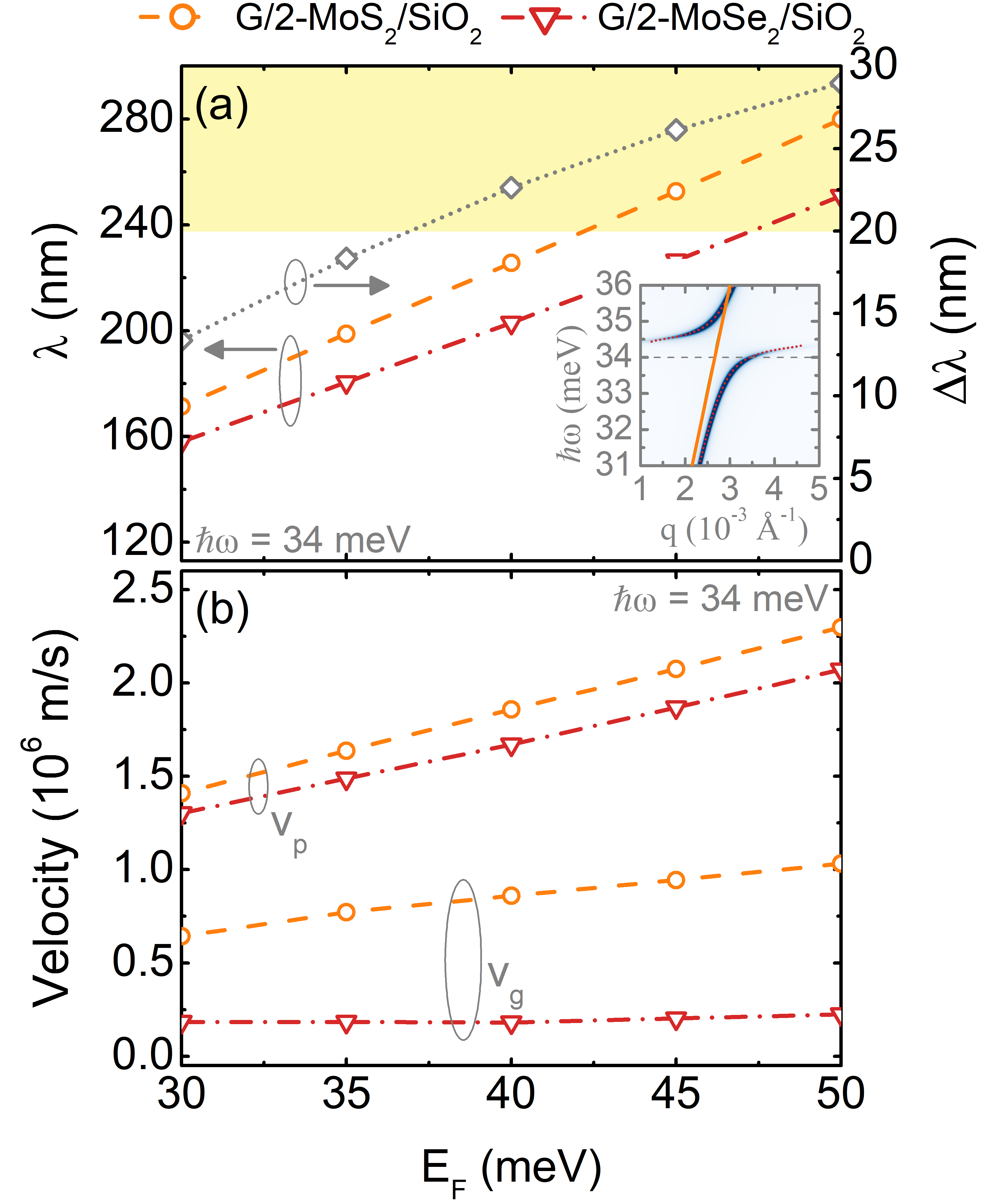}
\caption{(Color online) (a) Graphene surface plasmon-polariton wavelength as a function of the Fermi energy $E_{F}$ for $\text{G/2-MoS}_{2}\text{/SiO}_{2}$ (orange dashed line) and $\text{G/2-MoSe}_{2}\text{/SiO}_{2}$ (red dash-dotted line) and their respective differences $\Delta\lambda$ (gray dotted lines), at $\hbar\omega=34\text{ meV}$ ($\approx$\hspace{0.5mm}8.2~THz). The yellow region represents $\Delta\lambda\protect\geq20\text{ nm}$ (see right scale). The inset shows the plasmon-phonon dispersion at $E_{F}=45\text{ meV}$ near the hybridization region. The grey dashed line in the inset is at $\hbar\omega=34\text{ meV}$ and the orange solid line (dotted red lines) is the result for $\text{G/2-MoS}_{2}\text{/SiO}_{2}$ ($\text{G/2-MoSe}_{2}\text{/SiO}_{2}$). (b) Group velocity (\textsf{V}$_{\text{g}}$) and phase velocity (\textsf{V}$_{\text{p}}$) for the same vdWh as in (a).}
\label{Fig: Results_velocities}
\end{figure}

The bottom row of Fig.~\ref{Fig: Dispersion_MLG_MX2_ALL} shows the loss function when a SiO$_{2}$ substrate is added. A gray dashed horizontal line indicates the presence of the substrate surface phonons as discussed in the previous section. These substrate phonons also hybridize with the Dirac plasmons and render the high-frequency response of the different vdWhs almost identical. For lower frequencies, the additional SiO$_{2}$ environment significantly reduces the plasmon wavelength. But furthermore, the most significant effect of the presence of the substrate is the broadening of the A$_{2}''$ phonon spectrum.
A more in-depth analysis of the relation between the substrate phonons in SiO$_{2}$ and the TMD phonons is referred to future research. 

Finally, we are now in a position to propose a method to discriminate between different vdWhs based on the Dirac plasmon wavelength. To do so, in Fig.~\ref{Fig: Results_velocities}, we have calculated the wavelength of two types of heterostructures that differ only in one type of atom, $\rm MoS_{2}$ and $\rm MoSe_{2}$, at a given plasmon energy $\hbar \omega = 34~{\rm meV}$ ($\approx$\hspace{0.5mm}8.2~THz). In Fig.~\ref{Fig: Results_velocities}(a), the left axis gives the total value of the wavelength for given vdWhs, while the right axis refers to the difference $\Delta \lambda$ between both modes' wavelengths. The result indicates that once the Fermi level $E_{\rm F}$ is large enough, in the present case larger than 35 meV, $\Delta \lambda$ is large enough to be distinguished by current techniques~\citep{Woessner2014,Fei2012,Fei2015,Lundeberg2017,Lundeberg2016,Dai2015,Chen2012,Dai2014}. Notice that the difference between both heterostructures wavelengths depends also sensitively on the number of TMD layers in the vdWh. In the presented result, we assumed at least two TMD layers. For an increasing number of layers, the effect will be even stronger, rendering $N=2$ as the lower threshold for distinguishing between chemical components, which is a remarkably sensitive result. Fig.~\ref{Fig: Results_velocities}(b) shows the group (\textsf{V}$_{\text{g}}$) and phase velocity (\textsf{V}$_{\text{p}}$)  of the corresponding modes. Notice that, in this case, the group velocity of the $\rm MoSe_{2}$ heterostructure Dirac plasmon is almost constant as a function of the Fermi level. This is in stark contrast with the doping dependency of the $\rm MoS_{2}$-based system.

\section{Conclusions}\label{sec:conclusion}

We have demonstrated the possibility of using graphene plasmons to probe the non-local dynamical response of vdWHs composed by monolayer graphene on top of TMD multi-layers. In order to do so, we have calculated the loss function of graphene on top of different TMDs and demonstrated how its plasmon dispersion changes by the presence of the different materials and different numbers of layers underneath it. We have shown that the QEH model for this system provides excellent results when compared not only to available experimental data, but also to results obtained within the random phase approximation, rendering the QEH a good alternative for the theoretical understanding of experimental results involving plasmons in graphene-based vdWHs, as well as for the theoretical predictions shown here. Plasmons in graphene can be experimentally observed using, for example, scattering-type scanning near-field optical microscope (s-SNOM) in photocurrent mode, which has spatial resolution of at least 20 nm. Within this resolution, our results for the four TMD used here, namely $\text{MoS}_{2}$, $\text{MoSe}_{2}$, $\text{WS}_{2}$ and  $\text{WSe}_{2}$, show that it is possible to use surface plasmon-polaritons in the graphene monolayer to probe the number of layers in the TMD stack underneath it, by analyzing the difference in the plasmon wavelength as the number of layers change. Moreover, since different TMDs exhibit distinct phonon frequencies, the hybrid surface plasmon-phonon-polariton states can be used to identify which species of TMD is underneath the graphene layer. The latter, however, require strong coupling between plasmons and phonons to allow one to distinguish between the characteristic wavelengths of different TMDs. Nevertheless, our results show that for a number of layers as low as $N = 2$, the plasmon-phonon coupling is still strong enough to produce distinguishable wavelengths for different materials, thus suggesting the method proposed here as a remarkably sensitive tool.

\section*{ACKNOWLEDGMENTS}

This work was financially supported by the Brazilian Council for Research (CNPq), Brazilian National Council for the Improvement of Higher Education (CAPES) and by the Research Foundation Flanders (FWO) through a postdoctoral fellowship to B.V.D.

\appendix*
\section{Phononic structure of considered vdWh stacks }
\label{sec:Appendix:-phonons-frequencies}
Here we provide some important data for hBN and the four TMDs used in the main text.

\subsection{hBN reststrahlen bands}

It is important to mention that the upper and lower RS bands, shown in Fig.~\ref{Fig: Results_SP2_hBN_experimental}(b), obtained from the QEH, presents a small shift with respect to those obtained purely from first principles~\citep{Cai2007} and used as a reference in the two considered experiments~\citep{Woessner2014,Dai2015}. These values are provided in Tab.~\ref{Tab: RS_hBN}. There are, however, no qualitative differences, as shown in Fig.~\ref{Fig: Results_SP2_hBN_experimental}(e). 

The difference for the hyperbolic region II is highlighted by the horizontal gray dashed dotted-dotted lines (first principles~\citep{Cai2007}) and the dashed gray lines (QEH) in Fig.~\ref{Fig: Results_SP2_hBN_experimental}(e). However, the observed $\text{SP}^{3}$ and $\text{HP}^{3}$ modes, obtained from the QEH and the experimental methods, are in good agreement with each other, as compared in Figs.~\ref{Fig: Results_SP2_hBN_experimental}(d)-(e).

Table~\ref{Tab: RS_hBN} gives the phonon energies that define the two RS bands in hBN obtained from the QEH and those from first principles\citep{Cai2007}. As can be seen, these two methods differ in the order of $12\text{\%}$ ($1\sim2\text{\%}$) in the RS(I) (RS(II)). 

\begin{table}[!b]
\centering{}\caption{Frequencies that define the two Reststrahlen (RS) bands in hBN obtained from first principles calculations~\citep{Cai2007} and from the QEH model.}
\begin{tabular*}{1\columnwidth}{@{\extracolsep{\fill}}>{\centering}m{0.12\textwidth}>{\centering}m{0.08\textwidth}>{\centering}m{0.08\textwidth}>{\centering}m{0.08\textwidth}>{\centering}m{0.08\textwidth}}
 &  &  &  & \tabularnewline
\hline 
\hline 
\centering{} & \multicolumn{2}{c}{RS (I)} & \multicolumn{2}{c}{RS (II)}\tabularnewline
\hline 
 & $\hbar\omega_{TO}$ (meV) & $\hbar\omega_{LO}$ (meV) & $\hbar\omega_{TO}$ (meV) & $\hbar\omega_{LO}$ (meV)\tabularnewline
\hline 
First Prin.~\citep{Cai2007} & 96.70 & 102.90 & 169.85 & 199.61
\tabularnewline
QEH & 84.52 & 90.43 & 167.27 & 194.42\tabularnewline
Diferrence (\%) & 12.59 & 12.53 & 1.17 & 2.60\tabularnewline
\hline 
 &  &  &  & \tabularnewline
\end{tabular*}\label{Tab: RS_hBN}
\end{table}

\subsection{Phonon frequencies of the TMDs}

\begin{table}[!b]
\centering{}
\caption{Phonon frequencies for free-standing monolayer of $\text{MoS}_{2}$, $\text{WS}_{2}$, $\text{MoSe}_{2}$ and $\text{WSe}_{2}$ included in the QEH calculations. The relevant vibration modes are represent by $E''$, $E'$, $A'_{1}$ and $A''_{2}$.~ \cite{Zhang2015a,Zhao2013,MolinaSanchez2011,Peng2016,Berkdemir2013,Sengupta2015}}

\begin{tabular}{p{1cm} p{0.25cm} p{1.5cm} p{1.5cm} p{1.5cm} p{1.5cm}}
\hline\hline

&&   \multicolumn{4}{c}{Phonon frequencies (meV)}  \\ 
\hline
&& 1 ($E''$) & 2 ($E'$) & 3 ($A'_{1}$) & 4 ($A''_{2}$)\\
\hline
MoS$_{2}$ && 34.19 & 46.35 & 47.59 & 56.80 \\
\hline
WS$_{2}$ && 35.56 & 42.85 & 50.12 & 52.98 \\
\hline
MoSe$_{2}$ && 20.18 & 28.10 & 34.37 & 42.53 \\
\hline
WSe$_{2}$ && 20.71 & 29.67 & 30.19 & 37.21 \\
\hline\hline
\end{tabular}
\label{Tab: MX2-AppendixA}
\end{table}

All phonons frequencies for the transition metal dichalcogenide $\text{MX}_{2}$ ($\text{MoS}_{2}$, $\text{WS}_{2}$, $\text{MoSe}_{2}$ and $\text{WSe}_{2}$) included in the QEH calculations for $q\rightarrow0$, are provided in Tab.~\ref{Tab: MX2-AppendixA}. The optical modes of vibration are represented by $E''$, $E'$, $A'_{1}$ and $A''_{2}$ (for more details see Refs.~\cite{Zhang2015a,Zhao2013,MolinaSanchez2011,Peng2016,Berkdemir2013,Sengupta2015}).

\bibliographystyle{apsrev4-2}
\bibliography{myreferences}

\begin{thebibliography}{72}%
\makeatletter
\providecommand \@ifxundefined [1]{%
 \@ifx{#1\undefined}
}%
\providecommand \@ifnum [1]{%
 \ifnum #1\expandafter \@firstoftwo
 \else \expandafter \@secondoftwo
 \fi
}%
\providecommand \@ifx [1]{%
 \ifx #1\expandafter \@firstoftwo
 \else \expandafter \@secondoftwo
 \fi
}%
\providecommand \natexlab [1]{#1}%
\providecommand \enquote  [1]{``#1''}%
\providecommand \bibnamefont  [1]{#1}%
\providecommand \bibfnamefont [1]{#1}%
\providecommand \citenamefont [1]{#1}%
\providecommand \href@noop [0]{\@secondoftwo}%
\providecommand \href [0]{\begingroup \@sanitize@url \@href}%
\providecommand \@href[1]{\@@startlink{#1}\@@href}%
\providecommand \@@href[1]{\endgroup#1\@@endlink}%
\providecommand \@sanitize@url [0]{\catcode `\\12\catcode `\$12\catcode
  `\&12\catcode `\#12\catcode `\^12\catcode `\_12\catcode `\%12\relax}%
\providecommand \@@startlink[1]{}%
\providecommand \@@endlink[0]{}%
\providecommand \url  [0]{\begingroup\@sanitize@url \@url }%
\providecommand \@url [1]{\endgroup\@href {#1}{\urlprefix }}%
\providecommand \urlprefix  [0]{URL }%
\providecommand \Eprint [0]{\href }%
\providecommand \doibase [0]{https://doi.org/}%
\providecommand \selectlanguage [0]{\@gobble}%
\providecommand \bibinfo  [0]{\@secondoftwo}%
\providecommand \bibfield  [0]{\@secondoftwo}%
\providecommand \translation [1]{[#1]}%
\providecommand \BibitemOpen [0]{}%
\providecommand \bibitemStop [0]{}%
\providecommand \bibitemNoStop [0]{.\EOS\space}%
\providecommand \EOS [0]{\spacefactor3000\relax}%
\providecommand \BibitemShut  [1]{\csname bibitem#1\endcsname}%
\let\auto@bib@innerbib\@empty
\bibitem [{\citenamefont {Novoselov}(2004)}]{Novoselov2004}%
  \BibitemOpen
  \bibfield  {author} {\bibinfo {author} {\bibfnamefont {K.~S.}\ \bibnamefont
  {Novoselov}},\ }\href {https://doi.org/10.1126/science.1102896} {\bibfield
  {journal} {\bibinfo  {journal} {Science}\ }\textbf {\bibinfo {volume}
  {306}},\ \bibinfo {pages} {666} (\bibinfo {year} {2004})}\BibitemShut
  {NoStop}%
\bibitem [{\citenamefont {Manzeli}\ \emph {et~al.}(2017)\citenamefont
  {Manzeli}, \citenamefont {Ovchinnikov}, \citenamefont {Pasquier},
  \citenamefont {Yazyev},\ and\ \citenamefont {Kis}}]{Manzeli2017}%
  \BibitemOpen
  \bibfield  {author} {\bibinfo {author} {\bibfnamefont {S.}~\bibnamefont
  {Manzeli}}, \bibinfo {author} {\bibfnamefont {D.}~\bibnamefont
  {Ovchinnikov}}, \bibinfo {author} {\bibfnamefont {D.}~\bibnamefont
  {Pasquier}}, \bibinfo {author} {\bibfnamefont {O.~V.}\ \bibnamefont
  {Yazyev}},\ and\ \bibinfo {author} {\bibfnamefont {A.}~\bibnamefont {Kis}},\
  }\href {https://doi.org/10.1038/natrevmats.2017.33} {\bibfield  {journal}
  {\bibinfo  {journal} {Nature Reviews Materials}\ }\textbf {\bibinfo {volume}
  {2}},\ \bibinfo {pages} {17033} (\bibinfo {year} {2017})}\BibitemShut
  {NoStop}%
\bibitem [{\citenamefont {Novoselov}\ \emph {et~al.}(2005)\citenamefont
  {Novoselov}, \citenamefont {Jiang}, \citenamefont {Schedin}, \citenamefont
  {Booth}, \citenamefont {Khotkevich}, \citenamefont {Morozov},\ and\
  \citenamefont {Geim}}]{Novoselov2005}%
  \BibitemOpen
  \bibfield  {author} {\bibinfo {author} {\bibfnamefont {K.~S.}\ \bibnamefont
  {Novoselov}}, \bibinfo {author} {\bibfnamefont {D.}~\bibnamefont {Jiang}},
  \bibinfo {author} {\bibfnamefont {F.}~\bibnamefont {Schedin}}, \bibinfo
  {author} {\bibfnamefont {T.~J.}\ \bibnamefont {Booth}}, \bibinfo {author}
  {\bibfnamefont {V.~V.}\ \bibnamefont {Khotkevich}}, \bibinfo {author}
  {\bibfnamefont {S.~V.}\ \bibnamefont {Morozov}},\ and\ \bibinfo {author}
  {\bibfnamefont {A.~K.}\ \bibnamefont {Geim}},\ }\href
  {https://doi.org/10.1073/pnas.0502848102} {\bibfield  {journal} {\bibinfo
  {journal} {PNAS}\ }\textbf {\bibinfo {volume} {102}},\ \bibinfo {pages}
  {10451} (\bibinfo {year} {2005})}\BibitemShut {NoStop}%
\bibitem [{\citenamefont {Geim}\ and\ \citenamefont
  {Grigorieva}(2013)}]{Geim2013}%
  \BibitemOpen
  \bibfield  {author} {\bibinfo {author} {\bibfnamefont {A.~K.}\ \bibnamefont
  {Geim}}\ and\ \bibinfo {author} {\bibfnamefont {I.~V.}\ \bibnamefont
  {Grigorieva}},\ }\href {https://doi.org/10.1038/nature12385} {\bibfield
  {journal} {\bibinfo  {journal} {Nature}\ }\textbf {\bibinfo {volume} {499}},\
  \bibinfo {pages} {419} (\bibinfo {year} {2013})}\BibitemShut {NoStop}%
\bibitem [{\citenamefont {Wang}\ \emph {et~al.}(2012)\citenamefont {Wang},
  \citenamefont {Kalantar-Zadeh}, \citenamefont {Kis}, \citenamefont
  {Coleman},\ and\ \citenamefont {Strano}}]{Wang2012}%
  \BibitemOpen
  \bibfield  {author} {\bibinfo {author} {\bibfnamefont {Q.~H.}\ \bibnamefont
  {Wang}}, \bibinfo {author} {\bibfnamefont {K.}~\bibnamefont
  {Kalantar-Zadeh}}, \bibinfo {author} {\bibfnamefont {A.}~\bibnamefont {Kis}},
  \bibinfo {author} {\bibfnamefont {J.~N.}\ \bibnamefont {Coleman}},\ and\
  \bibinfo {author} {\bibfnamefont {M.~S.}\ \bibnamefont {Strano}},\ }\href
  {https://doi.org/10.1038/nnano.2012.193} {\bibfield  {journal} {\bibinfo
  {journal} {Nature Nanotechnology}\ }\textbf {\bibinfo {volume} {7}},\
  \bibinfo {pages} {699} (\bibinfo {year} {2012})}\BibitemShut {NoStop}%
\bibitem [{\citenamefont {Low}\ and\ \citenamefont {Avouris}(2014)}]{Low2014}%
  \BibitemOpen
  \bibfield  {author} {\bibinfo {author} {\bibfnamefont {T.}~\bibnamefont
  {Low}}\ and\ \bibinfo {author} {\bibfnamefont {P.}~\bibnamefont {Avouris}},\
  }\href {https://doi.org/10.1021/nn406627u} {\bibfield  {journal} {\bibinfo
  {journal} {{ACS} Nano}\ }\textbf {\bibinfo {volume} {8}},\ \bibinfo {pages}
  {1086} (\bibinfo {year} {2014})}\BibitemShut {NoStop}%
\bibitem [{\citenamefont {Low}\ \emph {et~al.}(2016)\citenamefont {Low},
  \citenamefont {Chaves}, \citenamefont {Caldwell}, \citenamefont {Kumar},
  \citenamefont {Fang}, \citenamefont {Avouris}, \citenamefont {Heinz},
  \citenamefont {Guinea}, \citenamefont {Martin-Moreno},\ and\ \citenamefont
  {Koppens}}]{Low2016}%
  \BibitemOpen
  \bibfield  {author} {\bibinfo {author} {\bibfnamefont {T.}~\bibnamefont
  {Low}}, \bibinfo {author} {\bibfnamefont {A.}~\bibnamefont {Chaves}},
  \bibinfo {author} {\bibfnamefont {J.~D.}\ \bibnamefont {Caldwell}}, \bibinfo
  {author} {\bibfnamefont {A.}~\bibnamefont {Kumar}}, \bibinfo {author}
  {\bibfnamefont {N.~X.}\ \bibnamefont {Fang}}, \bibinfo {author}
  {\bibfnamefont {P.}~\bibnamefont {Avouris}}, \bibinfo {author} {\bibfnamefont
  {T.~F.}\ \bibnamefont {Heinz}}, \bibinfo {author} {\bibfnamefont
  {F.}~\bibnamefont {Guinea}}, \bibinfo {author} {\bibfnamefont
  {L.}~\bibnamefont {Martin-Moreno}},\ and\ \bibinfo {author} {\bibfnamefont
  {F.}~\bibnamefont {Koppens}},\ }\href {https://doi.org/10.1038/nmat4792}
  {\bibfield  {journal} {\bibinfo  {journal} {Nature Materials}\ }\textbf
  {\bibinfo {volume} {16}},\ \bibinfo {pages} {182} (\bibinfo {year}
  {2016})}\BibitemShut {NoStop}%
\bibitem [{\citenamefont {Ranieri}(2014)}]{Ranieri2014}%
  \BibitemOpen
  \bibfield  {author} {\bibinfo {author} {\bibfnamefont {E.~D.}\ \bibnamefont
  {Ranieri}},\ }\href {https://doi.org/10.1038/nnano.2014.20} {\bibfield
  {journal} {\bibinfo  {journal} {Nature Nanotechnology}\ }\textbf {\bibinfo
  {volume} {5}},\ \bibinfo {pages} {3011} (\bibinfo {year} {2014})}\BibitemShut
  {NoStop}%
\bibitem [{\citenamefont {Jariwala}\ \emph {et~al.}(2014)\citenamefont
  {Jariwala}, \citenamefont {Sangwan}, \citenamefont {Lauhon}, \citenamefont
  {Marks},\ and\ \citenamefont {Hersam}}]{Jariwala2014}%
  \BibitemOpen
  \bibfield  {author} {\bibinfo {author} {\bibfnamefont {D.}~\bibnamefont
  {Jariwala}}, \bibinfo {author} {\bibfnamefont {V.~K.}\ \bibnamefont
  {Sangwan}}, \bibinfo {author} {\bibfnamefont {L.~J.}\ \bibnamefont {Lauhon}},
  \bibinfo {author} {\bibfnamefont {T.~J.}\ \bibnamefont {Marks}},\ and\
  \bibinfo {author} {\bibfnamefont {M.~C.}\ \bibnamefont {Hersam}},\ }\href
  {https://doi.org/10.1021/nn500064s} {\bibfield  {journal} {\bibinfo
  {journal} {{ACS} Nano}\ }\textbf {\bibinfo {volume} {8}},\ \bibinfo {pages}
  {1102} (\bibinfo {year} {2014})}\BibitemShut {NoStop}%
\bibitem [{\citenamefont {Zhang}(2015)}]{Zhang2015}%
  \BibitemOpen
  \bibfield  {author} {\bibinfo {author} {\bibfnamefont {H.}~\bibnamefont
  {Zhang}},\ }\href {https://doi.org/10.1021/acsnano.5b05040} {\bibfield
  {journal} {\bibinfo  {journal} {{ACS} Nano}\ }\textbf {\bibinfo {volume}
  {9}},\ \bibinfo {pages} {9451} (\bibinfo {year} {2015})}\BibitemShut
  {NoStop}%
\bibitem [{\citenamefont {Ju}\ \emph {et~al.}(2011)\citenamefont {Ju},
  \citenamefont {Geng}, \citenamefont {Horng}, \citenamefont {Girit},
  \citenamefont {Martin}, \citenamefont {Hao}, \citenamefont {Bechtel},
  \citenamefont {Liang}, \citenamefont {Zettl}, \citenamefont {Shen},\ and\
  \citenamefont {Wang}}]{Ju2011}%
  \BibitemOpen
  \bibfield  {author} {\bibinfo {author} {\bibfnamefont {L.}~\bibnamefont
  {Ju}}, \bibinfo {author} {\bibfnamefont {B.}~\bibnamefont {Geng}}, \bibinfo
  {author} {\bibfnamefont {J.}~\bibnamefont {Horng}}, \bibinfo {author}
  {\bibfnamefont {C.}~\bibnamefont {Girit}}, \bibinfo {author} {\bibfnamefont
  {M.}~\bibnamefont {Martin}}, \bibinfo {author} {\bibfnamefont
  {Z.}~\bibnamefont {Hao}}, \bibinfo {author} {\bibfnamefont {H.~A.}\
  \bibnamefont {Bechtel}}, \bibinfo {author} {\bibfnamefont {X.}~\bibnamefont
  {Liang}}, \bibinfo {author} {\bibfnamefont {A.}~\bibnamefont {Zettl}},
  \bibinfo {author} {\bibfnamefont {Y.~R.}\ \bibnamefont {Shen}},\ and\
  \bibinfo {author} {\bibfnamefont {F.}~\bibnamefont {Wang}},\ }\href
  {https://doi.org/10.1038/nnano.2011.146} {\bibfield  {journal} {\bibinfo
  {journal} {Nature Nanotechnology}\ }\textbf {\bibinfo {volume} {6}},\
  \bibinfo {pages} {630} (\bibinfo {year} {2011})}\BibitemShut {NoStop}%
\bibitem [{\citenamefont {Chen}\ \emph {et~al.}(2012)\citenamefont {Chen},
  \citenamefont {Badioli}, \citenamefont {Alonso-Gonz{\'{a}}lez}, \citenamefont
  {Thongrattanasiri}, \citenamefont {Huth}, \citenamefont {Osmond},
  \citenamefont {Spasenovi{\'{c}}}, \citenamefont {Centeno}, \citenamefont
  {Pesquera}, \citenamefont {Godignon}, \citenamefont {Elorza}, \citenamefont
  {Camara}, \citenamefont {de~Abajo}, \citenamefont {Hillenbrand},\ and\
  \citenamefont {Koppens}}]{Chen2012}%
  \BibitemOpen
  \bibfield  {author} {\bibinfo {author} {\bibfnamefont {J.}~\bibnamefont
  {Chen}}, \bibinfo {author} {\bibfnamefont {M.}~\bibnamefont {Badioli}},
  \bibinfo {author} {\bibfnamefont {P.}~\bibnamefont {Alonso-Gonz{\'{a}}lez}},
  \bibinfo {author} {\bibfnamefont {S.}~\bibnamefont {Thongrattanasiri}},
  \bibinfo {author} {\bibfnamefont {F.}~\bibnamefont {Huth}}, \bibinfo {author}
  {\bibfnamefont {J.}~\bibnamefont {Osmond}}, \bibinfo {author} {\bibfnamefont
  {M.}~\bibnamefont {Spasenovi{\'{c}}}}, \bibinfo {author} {\bibfnamefont
  {A.}~\bibnamefont {Centeno}}, \bibinfo {author} {\bibfnamefont
  {A.}~\bibnamefont {Pesquera}}, \bibinfo {author} {\bibfnamefont
  {P.}~\bibnamefont {Godignon}}, \bibinfo {author} {\bibfnamefont {A.~Z.}\
  \bibnamefont {Elorza}}, \bibinfo {author} {\bibfnamefont {N.}~\bibnamefont
  {Camara}}, \bibinfo {author} {\bibfnamefont {F.~J.~G.}\ \bibnamefont
  {de~Abajo}}, \bibinfo {author} {\bibfnamefont {R.}~\bibnamefont
  {Hillenbrand}},\ and\ \bibinfo {author} {\bibfnamefont {F.~H.~L.}\
  \bibnamefont {Koppens}},\ }\href {https://doi.org/10.1038/nature11254}
  {\bibfield  {journal} {\bibinfo  {journal} {Nature}\ }\textbf {\bibinfo
  {volume} {487}},\ \bibinfo {pages} {77} (\bibinfo {year} {2012})}\BibitemShut
  {NoStop}%
\bibitem [{\citenamefont {Fiori}\ \emph {et~al.}(2014)\citenamefont {Fiori},
  \citenamefont {Bonaccorso}, \citenamefont {Iannaccone}, \citenamefont
  {Palacios}, \citenamefont {Neumaier}, \citenamefont {Seabaugh}, \citenamefont
  {Banerjee},\ and\ \citenamefont {Colombo}}]{Fiori2014}%
  \BibitemOpen
  \bibfield  {author} {\bibinfo {author} {\bibfnamefont {G.}~\bibnamefont
  {Fiori}}, \bibinfo {author} {\bibfnamefont {F.}~\bibnamefont {Bonaccorso}},
  \bibinfo {author} {\bibfnamefont {G.}~\bibnamefont {Iannaccone}}, \bibinfo
  {author} {\bibfnamefont {T.}~\bibnamefont {Palacios}}, \bibinfo {author}
  {\bibfnamefont {D.}~\bibnamefont {Neumaier}}, \bibinfo {author}
  {\bibfnamefont {A.}~\bibnamefont {Seabaugh}}, \bibinfo {author}
  {\bibfnamefont {S.~K.}\ \bibnamefont {Banerjee}},\ and\ \bibinfo {author}
  {\bibfnamefont {L.}~\bibnamefont {Colombo}},\ }\href
  {https://doi.org/10.1038/nnano.2014.207} {\bibfield  {journal} {\bibinfo
  {journal} {Nature Nanotechnology}\ }\textbf {\bibinfo {volume} {9}},\
  \bibinfo {pages} {768} (\bibinfo {year} {2014})}\BibitemShut {NoStop}%
\bibitem [{\citenamefont {Mak}\ and\ \citenamefont {Shan}(2016)}]{Mak2016}%
  \BibitemOpen
  \bibfield  {author} {\bibinfo {author} {\bibfnamefont {K.~F.}\ \bibnamefont
  {Mak}}\ and\ \bibinfo {author} {\bibfnamefont {J.}~\bibnamefont {Shan}},\
  }\href {https://doi.org/10.1038/nphoton.2015.282} {\bibfield  {journal}
  {\bibinfo  {journal} {Nature Photonics}\ }\textbf {\bibinfo {volume} {10}},\
  \bibinfo {pages} {216} (\bibinfo {year} {2016})}\BibitemShut {NoStop}%
\bibitem [{\citenamefont {Yu}\ \emph {et~al.}(2015)\citenamefont {Yu},
  \citenamefont {Cai},\ and\ \citenamefont {Zhang}}]{Yu2015}%
  \BibitemOpen
  \bibfield  {author} {\bibinfo {author} {\bibfnamefont {Z.~G.}\ \bibnamefont
  {Yu}}, \bibinfo {author} {\bibfnamefont {Y.}~\bibnamefont {Cai}},\ and\
  \bibinfo {author} {\bibfnamefont {Y.-W.}\ \bibnamefont {Zhang}},\ }\href
  {https://doi.org/10.1038/srep13783} {\bibfield  {journal} {\bibinfo
  {journal} {Scientific Reports}\ }\textbf {\bibinfo {volume} {5}},\ \bibinfo
  {pages} {13783} (\bibinfo {year} {2015})}\BibitemShut {NoStop}%
\bibitem [{\citenamefont {Mak}\ \emph {et~al.}(2010)\citenamefont {Mak},
  \citenamefont {Lee}, \citenamefont {Hone}, \citenamefont {Shan},\ and\
  \citenamefont {Heinz}}]{Mak2010}%
  \BibitemOpen
  \bibfield  {author} {\bibinfo {author} {\bibfnamefont {K.~F.}\ \bibnamefont
  {Mak}}, \bibinfo {author} {\bibfnamefont {C.}~\bibnamefont {Lee}}, \bibinfo
  {author} {\bibfnamefont {J.}~\bibnamefont {Hone}}, \bibinfo {author}
  {\bibfnamefont {J.}~\bibnamefont {Shan}},\ and\ \bibinfo {author}
  {\bibfnamefont {T.~F.}\ \bibnamefont {Heinz}},\ }\href
  {https://doi.org/10.1103/physrevlett.105.136805} {\bibfield  {journal}
  {\bibinfo  {journal} {Physical Review Letters}\ }\textbf {\bibinfo {volume}
  {105}},\ \bibinfo {pages} {136805} (\bibinfo {year} {2010})}\BibitemShut
  {NoStop}%
\bibitem [{\citenamefont {Neto}\ \emph {et~al.}(2009)\citenamefont {Neto},
  \citenamefont {Guinea}, \citenamefont {Peres}, \citenamefont {Novoselov},\
  and\ \citenamefont {Geim}}]{Neto2009}%
  \BibitemOpen
  \bibfield  {author} {\bibinfo {author} {\bibfnamefont {A.~H.~C.}\
  \bibnamefont {Neto}}, \bibinfo {author} {\bibfnamefont {F.}~\bibnamefont
  {Guinea}}, \bibinfo {author} {\bibfnamefont {N.~M.~R.}\ \bibnamefont
  {Peres}}, \bibinfo {author} {\bibfnamefont {K.~S.}\ \bibnamefont
  {Novoselov}},\ and\ \bibinfo {author} {\bibfnamefont {A.~K.}\ \bibnamefont
  {Geim}},\ }\href {https://doi.org/10.1103/revmodphys.81.109} {\bibfield
  {journal} {\bibinfo  {journal} {Reviews of Modern Physics}\ }\textbf
  {\bibinfo {volume} {81}},\ \bibinfo {pages} {109} (\bibinfo {year}
  {2009})}\BibitemShut {NoStop}%
\bibitem [{\citenamefont {Gabriele~Giuliani}(2008)}]{GabrieleGiuliani2008}%
  \BibitemOpen
  \bibfield  {author} {\bibinfo {author} {\bibfnamefont {G.~V.}\ \bibnamefont
  {Gabriele~Giuliani}},\ }\href
  {https://www.ebook.de/de/product/7361265/gabriele_giuliani_giovanni_vignale_quantum_theory_of_the_electron_liquid.html}
  {\emph {\bibinfo {title} {Quantum Theory of the Electron Liquid}}}\ (\bibinfo
   {publisher} {Cambridge University Press},\ \bibinfo {year}
  {2008})\BibitemShut {NoStop}%
\bibitem [{\citenamefont {Maier}(2007)}]{Maier2007}%
  \BibitemOpen
  \bibfield  {author} {\bibinfo {author} {\bibfnamefont {S.~A.}\ \bibnamefont
  {Maier}},\ }\href
  {https://www.ebook.de/de/product/11430847/stefan_a_maier_plasmonics_fundamentals_and_applications.html}
  {\emph {\bibinfo {title} {Plasmonics: Fundamentals and Applications}}}\
  (\bibinfo  {publisher} {Springer-Verlag GmbH},\ \bibinfo {year}
  {2007})\BibitemShut {NoStop}%
\bibitem [{\citenamefont {Grigorenko}\ \emph {et~al.}(2012)\citenamefont
  {Grigorenko}, \citenamefont {Polini},\ and\ \citenamefont
  {Novoselov}}]{Grigorenko2012}%
  \BibitemOpen
  \bibfield  {author} {\bibinfo {author} {\bibfnamefont {A.~N.}\ \bibnamefont
  {Grigorenko}}, \bibinfo {author} {\bibfnamefont {M.}~\bibnamefont {Polini}},\
  and\ \bibinfo {author} {\bibfnamefont {K.~S.}\ \bibnamefont {Novoselov}},\
  }\href {https://doi.org/10.1038/nphoton.2012.262} {\bibfield  {journal}
  {\bibinfo  {journal} {Nature Photonics}\ }\textbf {\bibinfo {volume} {6}},\
  \bibinfo {pages} {749} (\bibinfo {year} {2012})}\BibitemShut {NoStop}%
\bibitem [{\citenamefont {Zhong}\ \emph {et~al.}(2015)\citenamefont {Zhong},
  \citenamefont {Malagari}, \citenamefont {Hamilton},\ and\ \citenamefont
  {Wasserman}}]{Zhong2015}%
  \BibitemOpen
  \bibfield  {author} {\bibinfo {author} {\bibfnamefont {Y.}~\bibnamefont
  {Zhong}}, \bibinfo {author} {\bibfnamefont {S.~D.}\ \bibnamefont {Malagari}},
  \bibinfo {author} {\bibfnamefont {T.}~\bibnamefont {Hamilton}},\ and\
  \bibinfo {author} {\bibfnamefont {D.}~\bibnamefont {Wasserman}},\ }\href
  {https://doi.org/10.1117/1.jnp.9.093791} {\bibfield  {journal} {\bibinfo
  {journal} {J. Nanophotonics}\ }\textbf {\bibinfo {volume} {9}},\ \bibinfo
  {pages} {093791} (\bibinfo {year} {2015})}\BibitemShut {NoStop}%
\bibitem [{\citenamefont {Schuller}\ \emph {et~al.}(2010)\citenamefont
  {Schuller}, \citenamefont {Barnard}, \citenamefont {Cai}, \citenamefont
  {Jun}, \citenamefont {White},\ and\ \citenamefont
  {Brongersma}}]{Schuller2010}%
  \BibitemOpen
  \bibfield  {author} {\bibinfo {author} {\bibfnamefont {J.~A.}\ \bibnamefont
  {Schuller}}, \bibinfo {author} {\bibfnamefont {E.~S.}\ \bibnamefont
  {Barnard}}, \bibinfo {author} {\bibfnamefont {W.}~\bibnamefont {Cai}},
  \bibinfo {author} {\bibfnamefont {Y.~C.}\ \bibnamefont {Jun}}, \bibinfo
  {author} {\bibfnamefont {J.~S.}\ \bibnamefont {White}},\ and\ \bibinfo
  {author} {\bibfnamefont {M.~L.}\ \bibnamefont {Brongersma}},\ }\href
  {https://doi.org/10.1038/nmat2630} {\bibfield  {journal} {\bibinfo  {journal}
  {Nature Materials}\ }\textbf {\bibinfo {volume} {9}},\ \bibinfo {pages} {193}
  (\bibinfo {year} {2010})}\BibitemShut {NoStop}%
\bibitem [{\citenamefont {Polini}(2016)}]{Polini2016}%
  \BibitemOpen
  \bibfield  {author} {\bibinfo {author} {\bibfnamefont {M.}~\bibnamefont
  {Polini}},\ }\href {https://doi.org/10.1126/science.aad7995} {\bibfield
  {journal} {\bibinfo  {journal} {Science}\ }\textbf {\bibinfo {volume}
  {351}},\ \bibinfo {pages} {229} (\bibinfo {year} {2016})}\BibitemShut
  {NoStop}%
\bibitem [{\citenamefont {Alonso-Gonz{\'a}lez}\ \emph
  {et~al.}(2016)\citenamefont {Alonso-Gonz{\'a}lez}, \citenamefont {Nikitin},
  \citenamefont {Gao}, \citenamefont {Woessner}, \citenamefont {Lundeberg},
  \citenamefont {Principi}, \citenamefont {Forcellini}, \citenamefont {Yan},
  \citenamefont {V{\'e}lez}, \citenamefont {Huber} \emph
  {et~al.}}]{Alonso-Gonzalez2016}%
  \BibitemOpen
  \bibfield  {author} {\bibinfo {author} {\bibfnamefont {P.}~\bibnamefont
  {Alonso-Gonz{\'a}lez}}, \bibinfo {author} {\bibfnamefont {A.~Y.}\
  \bibnamefont {Nikitin}}, \bibinfo {author} {\bibfnamefont {Y.}~\bibnamefont
  {Gao}}, \bibinfo {author} {\bibfnamefont {A.}~\bibnamefont {Woessner}},
  \bibinfo {author} {\bibfnamefont {M.~B.}\ \bibnamefont {Lundeberg}}, \bibinfo
  {author} {\bibfnamefont {A.}~\bibnamefont {Principi}}, \bibinfo {author}
  {\bibfnamefont {N.}~\bibnamefont {Forcellini}}, \bibinfo {author}
  {\bibfnamefont {W.}~\bibnamefont {Yan}}, \bibinfo {author} {\bibfnamefont
  {S.}~\bibnamefont {V{\'e}lez}}, \bibinfo {author} {\bibfnamefont {A.~J.}\
  \bibnamefont {Huber}}, \emph {et~al.},\ }\href
  {https://doi.org/10.1038/nnano.2016.185} {\bibfield  {journal} {\bibinfo
  {journal} {Nature Nanotechnology}\ }\textbf {\bibinfo {volume} {12}},\
  \bibinfo {pages} {31} (\bibinfo {year} {2016})}\BibitemShut {NoStop}%
\bibitem [{\citenamefont {Gon{\c{c}}alves}\ and\ \citenamefont
  {Peres}(2015)}]{Goncalves2015}%
  \BibitemOpen
  \bibfield  {author} {\bibinfo {author} {\bibfnamefont {P.~A.~D.}\
  \bibnamefont {Gon{\c{c}}alves}}\ and\ \bibinfo {author} {\bibfnamefont
  {N.~M.~R.}\ \bibnamefont {Peres}},\ }\href {https://doi.org/10.1142/9948}
  {\emph {\bibinfo {title} {An Introduction to Graphene Plasmonics}}}\
  (\bibinfo  {publisher} {{World} {Scienctific}},\ \bibinfo {year}
  {2015})\BibitemShut {NoStop}%
\bibitem [{\citenamefont {Splendiani}\ \emph {et~al.}(2010)\citenamefont
  {Splendiani}, \citenamefont {Sun}, \citenamefont {Zhang}, \citenamefont {Li},
  \citenamefont {Kim}, \citenamefont {Chim}, \citenamefont {Galli},\ and\
  \citenamefont {Wang}}]{Splendiani2010}%
  \BibitemOpen
  \bibfield  {author} {\bibinfo {author} {\bibfnamefont {A.}~\bibnamefont
  {Splendiani}}, \bibinfo {author} {\bibfnamefont {L.}~\bibnamefont {Sun}},
  \bibinfo {author} {\bibfnamefont {Y.}~\bibnamefont {Zhang}}, \bibinfo
  {author} {\bibfnamefont {T.}~\bibnamefont {Li}}, \bibinfo {author}
  {\bibfnamefont {J.}~\bibnamefont {Kim}}, \bibinfo {author} {\bibfnamefont
  {C.-Y.}\ \bibnamefont {Chim}}, \bibinfo {author} {\bibfnamefont
  {G.}~\bibnamefont {Galli}},\ and\ \bibinfo {author} {\bibfnamefont
  {F.}~\bibnamefont {Wang}},\ }\href {https://doi.org/10.1021/nl903868w}
  {\bibfield  {journal} {\bibinfo  {journal} {Nano Letters}\ }\textbf {\bibinfo
  {volume} {10}},\ \bibinfo {pages} {1271} (\bibinfo {year}
  {2010})}\BibitemShut {NoStop}%
\bibitem [{\citenamefont {Li}\ \emph {et~al.}(2015)\citenamefont {Li},
  \citenamefont {Xiao}, \citenamefont {Gong}, \citenamefont {Wang},
  \citenamefont {Kang}, \citenamefont {Zu}, \citenamefont {Ajayan},
  \citenamefont {Nordlander},\ and\ \citenamefont {Fang}}]{Li2015}%
  \BibitemOpen
  \bibfield  {author} {\bibinfo {author} {\bibfnamefont {Z.}~\bibnamefont
  {Li}}, \bibinfo {author} {\bibfnamefont {Y.}~\bibnamefont {Xiao}}, \bibinfo
  {author} {\bibfnamefont {Y.}~\bibnamefont {Gong}}, \bibinfo {author}
  {\bibfnamefont {Z.}~\bibnamefont {Wang}}, \bibinfo {author} {\bibfnamefont
  {Y.}~\bibnamefont {Kang}}, \bibinfo {author} {\bibfnamefont {S.}~\bibnamefont
  {Zu}}, \bibinfo {author} {\bibfnamefont {P.~M.}\ \bibnamefont {Ajayan}},
  \bibinfo {author} {\bibfnamefont {P.}~\bibnamefont {Nordlander}},\ and\
  \bibinfo {author} {\bibfnamefont {Z.}~\bibnamefont {Fang}},\ }\href
  {https://doi.org/10.1021/acsnano.5b03764} {\bibfield  {journal} {\bibinfo
  {journal} {{ACS} Nano}\ }\textbf {\bibinfo {volume} {9}},\ \bibinfo {pages}
  {10158} (\bibinfo {year} {2015})}\BibitemShut {NoStop}%
\bibitem [{\citenamefont {Gong}\ \emph {et~al.}(2014)\citenamefont {Gong},
  \citenamefont {Lin}, \citenamefont {Wang}, \citenamefont {Shi}, \citenamefont
  {Lei}, \citenamefont {Lin}, \citenamefont {Zou}, \citenamefont {Ye},
  \citenamefont {Vajtai}, \citenamefont {Yakobson} \emph
  {et~al.}}]{gong2014vertical}%
  \BibitemOpen
  \bibfield  {author} {\bibinfo {author} {\bibfnamefont {Y.}~\bibnamefont
  {Gong}}, \bibinfo {author} {\bibfnamefont {J.}~\bibnamefont {Lin}}, \bibinfo
  {author} {\bibfnamefont {X.}~\bibnamefont {Wang}}, \bibinfo {author}
  {\bibfnamefont {G.}~\bibnamefont {Shi}}, \bibinfo {author} {\bibfnamefont
  {S.}~\bibnamefont {Lei}}, \bibinfo {author} {\bibfnamefont {Z.}~\bibnamefont
  {Lin}}, \bibinfo {author} {\bibfnamefont {X.}~\bibnamefont {Zou}}, \bibinfo
  {author} {\bibfnamefont {G.}~\bibnamefont {Ye}}, \bibinfo {author}
  {\bibfnamefont {R.}~\bibnamefont {Vajtai}}, \bibinfo {author} {\bibfnamefont
  {B.~I.}\ \bibnamefont {Yakobson}}, \emph {et~al.},\ }\href
  {https://doi.org/https://doi.org/10.1038/nmat4091} {\bibfield  {journal}
  {\bibinfo  {journal} {Nature materials}\ }\textbf {\bibinfo {volume} {13}},\
  \bibinfo {pages} {1135} (\bibinfo {year} {2014})}\BibitemShut {NoStop}%
\bibitem [{\citenamefont {{\"O}zcelik}\ \emph {et~al.}(2016)\citenamefont
  {{\"O}zcelik}, \citenamefont {Azadani}, \citenamefont {Yang}, \citenamefont
  {Koester},\ and\ \citenamefont {Low}}]{ozcelik2016band}%
  \BibitemOpen
  \bibfield  {author} {\bibinfo {author} {\bibfnamefont {V.~O.}\ \bibnamefont
  {{\"O}zcelik}}, \bibinfo {author} {\bibfnamefont {J.~G.}\ \bibnamefont
  {Azadani}}, \bibinfo {author} {\bibfnamefont {C.}~\bibnamefont {Yang}},
  \bibinfo {author} {\bibfnamefont {S.~J.}\ \bibnamefont {Koester}},\ and\
  \bibinfo {author} {\bibfnamefont {T.}~\bibnamefont {Low}},\ }\href
  {https://doi.org/https://doi.org/10.1103/PhysRevB.94.035125} {\bibfield
  {journal} {\bibinfo  {journal} {Physical Review B}\ }\textbf {\bibinfo
  {volume} {94}},\ \bibinfo {pages} {035125} (\bibinfo {year}
  {2016})}\BibitemShut {NoStop}%
\bibitem [{\citenamefont {Sahoo}\ \emph {et~al.}(2018)\citenamefont {Sahoo},
  \citenamefont {Memaran}, \citenamefont {Xin}, \citenamefont {Balicas},\ and\
  \citenamefont {Guti{\'e}rrez}}]{sahoo2018one}%
  \BibitemOpen
  \bibfield  {author} {\bibinfo {author} {\bibfnamefont {P.~K.}\ \bibnamefont
  {Sahoo}}, \bibinfo {author} {\bibfnamefont {S.}~\bibnamefont {Memaran}},
  \bibinfo {author} {\bibfnamefont {Y.}~\bibnamefont {Xin}}, \bibinfo {author}
  {\bibfnamefont {L.}~\bibnamefont {Balicas}},\ and\ \bibinfo {author}
  {\bibfnamefont {H.~R.}\ \bibnamefont {Guti{\'e}rrez}},\ }\href
  {https://doi.org/https://doi.org/10.1038/nature25155} {\bibfield  {journal}
  {\bibinfo  {journal} {Nature}\ }\textbf {\bibinfo {volume} {553}},\ \bibinfo
  {pages} {63} (\bibinfo {year} {2018})}\BibitemShut {NoStop}%
\bibitem [{\citenamefont {Duan}\ \emph {et~al.}(2014)\citenamefont {Duan},
  \citenamefont {Wang}, \citenamefont {Shaw}, \citenamefont {Cheng},
  \citenamefont {Chen}, \citenamefont {Li}, \citenamefont {Wu}, \citenamefont
  {Tang}, \citenamefont {Zhang}, \citenamefont {Pan} \emph
  {et~al.}}]{duan2014lateral}%
  \BibitemOpen
  \bibfield  {author} {\bibinfo {author} {\bibfnamefont {X.}~\bibnamefont
  {Duan}}, \bibinfo {author} {\bibfnamefont {C.}~\bibnamefont {Wang}}, \bibinfo
  {author} {\bibfnamefont {J.~C.}\ \bibnamefont {Shaw}}, \bibinfo {author}
  {\bibfnamefont {R.}~\bibnamefont {Cheng}}, \bibinfo {author} {\bibfnamefont
  {Y.}~\bibnamefont {Chen}}, \bibinfo {author} {\bibfnamefont {H.}~\bibnamefont
  {Li}}, \bibinfo {author} {\bibfnamefont {X.}~\bibnamefont {Wu}}, \bibinfo
  {author} {\bibfnamefont {Y.}~\bibnamefont {Tang}}, \bibinfo {author}
  {\bibfnamefont {Q.}~\bibnamefont {Zhang}}, \bibinfo {author} {\bibfnamefont
  {A.}~\bibnamefont {Pan}}, \emph {et~al.},\ }\href
  {https://doi.org/https://doi.org/10.1038/nnano.2014.222} {\bibfield
  {journal} {\bibinfo  {journal} {Nature nanotechnology}\ }\textbf {\bibinfo
  {volume} {9}},\ \bibinfo {pages} {1024} (\bibinfo {year} {2014})}\BibitemShut
  {NoStop}%
\bibitem [{\citenamefont {Gong}\ \emph {et~al.}(2015)\citenamefont {Gong},
  \citenamefont {Lei}, \citenamefont {Ye}, \citenamefont {Li}, \citenamefont
  {He}, \citenamefont {Keyshar}, \citenamefont {Zhang}, \citenamefont {Wang},
  \citenamefont {Lou}, \citenamefont {Liu} \emph {et~al.}}]{gong2015two}%
  \BibitemOpen
  \bibfield  {author} {\bibinfo {author} {\bibfnamefont {Y.}~\bibnamefont
  {Gong}}, \bibinfo {author} {\bibfnamefont {S.}~\bibnamefont {Lei}}, \bibinfo
  {author} {\bibfnamefont {G.}~\bibnamefont {Ye}}, \bibinfo {author}
  {\bibfnamefont {B.}~\bibnamefont {Li}}, \bibinfo {author} {\bibfnamefont
  {Y.}~\bibnamefont {He}}, \bibinfo {author} {\bibfnamefont {K.}~\bibnamefont
  {Keyshar}}, \bibinfo {author} {\bibfnamefont {X.}~\bibnamefont {Zhang}},
  \bibinfo {author} {\bibfnamefont {Q.}~\bibnamefont {Wang}}, \bibinfo {author}
  {\bibfnamefont {J.}~\bibnamefont {Lou}}, \bibinfo {author} {\bibfnamefont
  {Z.}~\bibnamefont {Liu}}, \emph {et~al.},\ }\href
  {https://doi.org/https://doi.org/10.1021/acs.nanolett.5b02423} {\bibfield
  {journal} {\bibinfo  {journal} {Nano letters}\ }\textbf {\bibinfo {volume}
  {15}},\ \bibinfo {pages} {6135} (\bibinfo {year} {2015})}\BibitemShut
  {NoStop}%
\bibitem [{\citenamefont {Huang}\ \emph {et~al.}(2014)\citenamefont {Huang},
  \citenamefont {Wu}, \citenamefont {Sanchez}, \citenamefont {Peters},
  \citenamefont {Beanland}, \citenamefont {Ross}, \citenamefont {Rivera},
  \citenamefont {Yao}, \citenamefont {Cobden},\ and\ \citenamefont
  {Xu}}]{huang2014lateral}%
  \BibitemOpen
  \bibfield  {author} {\bibinfo {author} {\bibfnamefont {C.}~\bibnamefont
  {Huang}}, \bibinfo {author} {\bibfnamefont {S.}~\bibnamefont {Wu}}, \bibinfo
  {author} {\bibfnamefont {A.~M.}\ \bibnamefont {Sanchez}}, \bibinfo {author}
  {\bibfnamefont {J.~J.}\ \bibnamefont {Peters}}, \bibinfo {author}
  {\bibfnamefont {R.}~\bibnamefont {Beanland}}, \bibinfo {author}
  {\bibfnamefont {J.~S.}\ \bibnamefont {Ross}}, \bibinfo {author}
  {\bibfnamefont {P.}~\bibnamefont {Rivera}}, \bibinfo {author} {\bibfnamefont
  {W.}~\bibnamefont {Yao}}, \bibinfo {author} {\bibfnamefont {D.~H.}\
  \bibnamefont {Cobden}},\ and\ \bibinfo {author} {\bibfnamefont
  {X.}~\bibnamefont {Xu}},\ }\href
  {https://doi.org/https://doi.org/10.1038/nmat4064} {\bibfield  {journal}
  {\bibinfo  {journal} {Nature materials}\ }\textbf {\bibinfo {volume} {13}},\
  \bibinfo {pages} {1096} (\bibinfo {year} {2014})}\BibitemShut {NoStop}%
\bibitem [{\citenamefont {Jariwala}\ \emph {et~al.}(2016)\citenamefont
  {Jariwala}, \citenamefont {Marks},\ and\ \citenamefont
  {Hersam}}]{Jariwala2016}%
  \BibitemOpen
  \bibfield  {author} {\bibinfo {author} {\bibfnamefont {D.}~\bibnamefont
  {Jariwala}}, \bibinfo {author} {\bibfnamefont {T.~J.}\ \bibnamefont
  {Marks}},\ and\ \bibinfo {author} {\bibfnamefont {M.~C.}\ \bibnamefont
  {Hersam}},\ }\href {https://doi.org/10.1038/nmat4703} {\bibfield  {journal}
  {\bibinfo  {journal} {Nature Materials}\ }\textbf {\bibinfo {volume} {16}},\
  \bibinfo {pages} {170} (\bibinfo {year} {2016})}\BibitemShut {NoStop}%
\bibitem [{\citenamefont {Liu}\ \emph {et~al.}(2016)\citenamefont {Liu},
  \citenamefont {Weiss}, \citenamefont {Duan}, \citenamefont {Cheng},
  \citenamefont {Huang},\ and\ \citenamefont {Duan}}]{Liu2016}%
  \BibitemOpen
  \bibfield  {author} {\bibinfo {author} {\bibfnamefont {Y.}~\bibnamefont
  {Liu}}, \bibinfo {author} {\bibfnamefont {N.~O.}\ \bibnamefont {Weiss}},
  \bibinfo {author} {\bibfnamefont {X.}~\bibnamefont {Duan}}, \bibinfo {author}
  {\bibfnamefont {H.-C.}\ \bibnamefont {Cheng}}, \bibinfo {author}
  {\bibfnamefont {Y.}~\bibnamefont {Huang}},\ and\ \bibinfo {author}
  {\bibfnamefont {X.}~\bibnamefont {Duan}},\ }\href
  {https://doi.org/10.1038/natrevmats.2016.42} {\bibfield  {journal} {\bibinfo
  {journal} {Nature Reviews Materials}\ }\textbf {\bibinfo {volume} {1}},\
  \bibinfo {pages} {1} (\bibinfo {year} {2016})}\BibitemShut {NoStop}%
\bibitem [{\citenamefont {Chaves}\ \emph {et~al.}(2020)\citenamefont {Chaves},
  \citenamefont {Azadani}, \citenamefont {Alsalman}, \citenamefont {da~Costa},
  \citenamefont {Frisenda}, \citenamefont {Chaves}, \citenamefont {Song},
  \citenamefont {Kim}, \citenamefont {He}, \citenamefont {Zhou} \emph
  {et~al.}}]{chaves2020bandgap}%
  \BibitemOpen
  \bibfield  {author} {\bibinfo {author} {\bibfnamefont {A.}~\bibnamefont
  {Chaves}}, \bibinfo {author} {\bibfnamefont {J.}~\bibnamefont {Azadani}},
  \bibinfo {author} {\bibfnamefont {H.}~\bibnamefont {Alsalman}}, \bibinfo
  {author} {\bibfnamefont {D.}~\bibnamefont {da~Costa}}, \bibinfo {author}
  {\bibfnamefont {R.}~\bibnamefont {Frisenda}}, \bibinfo {author}
  {\bibfnamefont {A.}~\bibnamefont {Chaves}}, \bibinfo {author} {\bibfnamefont
  {S.~H.}\ \bibnamefont {Song}}, \bibinfo {author} {\bibfnamefont
  {Y.}~\bibnamefont {Kim}}, \bibinfo {author} {\bibfnamefont {D.}~\bibnamefont
  {He}}, \bibinfo {author} {\bibfnamefont {J.}~\bibnamefont {Zhou}}, \emph
  {et~al.},\ }\href
  {https://doi.org/https://doi.org/10.1038/s41699-020-00162-4} {\bibfield
  {journal} {\bibinfo  {journal} {npj 2D Materials and Applications}\ }\textbf
  {\bibinfo {volume} {4}},\ \bibinfo {pages} {1} (\bibinfo {year}
  {2020})}\BibitemShut {NoStop}%
\bibitem [{\citenamefont {Griffiths}(2017)}]{Griffiths2017}%
  \BibitemOpen
  \bibfield  {author} {\bibinfo {author} {\bibfnamefont {D.~J.}\ \bibnamefont
  {Griffiths}},\ }\href
  {https://www.ebook.de/de/product/29245261/david_j_griffiths_introduction_to_electrodynamics.html}
  {\emph {\bibinfo {title} {Introduction to Electrodynamics}}}\ (\bibinfo
  {publisher} {Cambridge University Press},\ \bibinfo {year}
  {2017})\BibitemShut {NoStop}%
\bibitem [{\citenamefont {Luxmoore}\ \emph {et~al.}(2014)\citenamefont
  {Luxmoore}, \citenamefont {Gan}, \citenamefont {Liu}, \citenamefont
  {Valmorra}, \citenamefont {Li}, \citenamefont {Faist},\ and\ \citenamefont
  {Nash}}]{Luxmoore2014}%
  \BibitemOpen
  \bibfield  {author} {\bibinfo {author} {\bibfnamefont {I.~J.}\ \bibnamefont
  {Luxmoore}}, \bibinfo {author} {\bibfnamefont {C.~H.}\ \bibnamefont {Gan}},
  \bibinfo {author} {\bibfnamefont {P.~Q.}\ \bibnamefont {Liu}}, \bibinfo
  {author} {\bibfnamefont {F.}~\bibnamefont {Valmorra}}, \bibinfo {author}
  {\bibfnamefont {P.}~\bibnamefont {Li}}, \bibinfo {author} {\bibfnamefont
  {J.}~\bibnamefont {Faist}},\ and\ \bibinfo {author} {\bibfnamefont {G.~R.}\
  \bibnamefont {Nash}},\ }\href {https://doi.org/10.1021/ph500233s} {\bibfield
  {journal} {\bibinfo  {journal} {{ACS} Photonics}\ }\textbf {\bibinfo {volume}
  {1}},\ \bibinfo {pages} {1151} (\bibinfo {year} {2014})}\BibitemShut
  {NoStop}%
\bibitem [{\citenamefont {Wu}\ \emph {et~al.}(2016)\citenamefont {Wu},
  \citenamefont {Jiang}, \citenamefont {Xu}, \citenamefont {Dai}, \citenamefont
  {Xiang},\ and\ \citenamefont {Fan}}]{Wu2016}%
  \BibitemOpen
  \bibfield  {author} {\bibinfo {author} {\bibfnamefont {Y.}~\bibnamefont
  {Wu}}, \bibinfo {author} {\bibfnamefont {L.}~\bibnamefont {Jiang}}, \bibinfo
  {author} {\bibfnamefont {H.}~\bibnamefont {Xu}}, \bibinfo {author}
  {\bibfnamefont {X.}~\bibnamefont {Dai}}, \bibinfo {author} {\bibfnamefont
  {Y.}~\bibnamefont {Xiang}},\ and\ \bibinfo {author} {\bibfnamefont
  {D.}~\bibnamefont {Fan}},\ }\href {https://doi.org/10.1364/oe.24.002109}
  {\bibfield  {journal} {\bibinfo  {journal} {Optics Express}\ }\textbf
  {\bibinfo {volume} {24}},\ \bibinfo {pages} {2109} (\bibinfo {year}
  {2016})}\BibitemShut {NoStop}%
\bibitem [{\citenamefont {Dai}\ \emph {et~al.}(2015)\citenamefont {Dai},
  \citenamefont {Ma}, \citenamefont {Liu}, \citenamefont {Andersen},
  \citenamefont {Fei}, \citenamefont {Goldflam}, \citenamefont {Wagner},
  \citenamefont {Watanabe}, \citenamefont {Taniguchi}, \citenamefont {Thiemens}
  \emph {et~al.}}]{Dai2015}%
  \BibitemOpen
  \bibfield  {author} {\bibinfo {author} {\bibfnamefont {S.}~\bibnamefont
  {Dai}}, \bibinfo {author} {\bibfnamefont {Q.}~\bibnamefont {Ma}}, \bibinfo
  {author} {\bibfnamefont {M.}~\bibnamefont {Liu}}, \bibinfo {author}
  {\bibfnamefont {T.}~\bibnamefont {Andersen}}, \bibinfo {author}
  {\bibfnamefont {Z.}~\bibnamefont {Fei}}, \bibinfo {author} {\bibfnamefont
  {M.}~\bibnamefont {Goldflam}}, \bibinfo {author} {\bibfnamefont
  {M.}~\bibnamefont {Wagner}}, \bibinfo {author} {\bibfnamefont
  {K.}~\bibnamefont {Watanabe}}, \bibinfo {author} {\bibfnamefont
  {T.}~\bibnamefont {Taniguchi}}, \bibinfo {author} {\bibfnamefont
  {M.}~\bibnamefont {Thiemens}}, \emph {et~al.},\ }\href
  {https://doi.org/10.1038/nnano.2015.131} {\bibfield  {journal} {\bibinfo
  {journal} {Nature Nanotechnology}\ }\textbf {\bibinfo {volume} {10}},\
  \bibinfo {pages} {682} (\bibinfo {year} {2015})}\BibitemShut {NoStop}%
\bibitem [{\citenamefont {Lundeberg}\ \emph {et~al.}(2017)\citenamefont
  {Lundeberg}, \citenamefont {Gao}, \citenamefont {Asgari}, \citenamefont
  {Tan}, \citenamefont {Van~Duppen}, \citenamefont {Autore}, \citenamefont
  {Alonso-Gonz{\'a}lez}, \citenamefont {Woessner}, \citenamefont {Watanabe},
  \citenamefont {Taniguchi} \emph {et~al.}}]{Lundeberg2017}%
  \BibitemOpen
  \bibfield  {author} {\bibinfo {author} {\bibfnamefont {M.~B.}\ \bibnamefont
  {Lundeberg}}, \bibinfo {author} {\bibfnamefont {Y.}~\bibnamefont {Gao}},
  \bibinfo {author} {\bibfnamefont {R.}~\bibnamefont {Asgari}}, \bibinfo
  {author} {\bibfnamefont {C.}~\bibnamefont {Tan}}, \bibinfo {author}
  {\bibfnamefont {B.}~\bibnamefont {Van~Duppen}}, \bibinfo {author}
  {\bibfnamefont {M.}~\bibnamefont {Autore}}, \bibinfo {author} {\bibfnamefont
  {P.}~\bibnamefont {Alonso-Gonz{\'a}lez}}, \bibinfo {author} {\bibfnamefont
  {A.}~\bibnamefont {Woessner}}, \bibinfo {author} {\bibfnamefont
  {K.}~\bibnamefont {Watanabe}}, \bibinfo {author} {\bibfnamefont
  {T.}~\bibnamefont {Taniguchi}}, \emph {et~al.},\ }\href
  {https://doi.org/10.1126/science.aan2735} {\bibfield  {journal} {\bibinfo
  {journal} {Science}\ }\textbf {\bibinfo {volume} {357}},\ \bibinfo {pages}
  {187} (\bibinfo {year} {2017})}\BibitemShut {NoStop}%
\bibitem [{\citenamefont {Lundeberg}\ \emph {et~al.}(2016)\citenamefont
  {Lundeberg}, \citenamefont {Gao}, \citenamefont {Woessner}, \citenamefont
  {Tan}, \citenamefont {Alonso-Gonz{\'a}lez}, \citenamefont {Watanabe},
  \citenamefont {Taniguchi}, \citenamefont {Hone}, \citenamefont
  {Hillenbrand},\ and\ \citenamefont {Koppens}}]{Lundeberg2016}%
  \BibitemOpen
  \bibfield  {author} {\bibinfo {author} {\bibfnamefont {M.~B.}\ \bibnamefont
  {Lundeberg}}, \bibinfo {author} {\bibfnamefont {Y.}~\bibnamefont {Gao}},
  \bibinfo {author} {\bibfnamefont {A.}~\bibnamefont {Woessner}}, \bibinfo
  {author} {\bibfnamefont {C.}~\bibnamefont {Tan}}, \bibinfo {author}
  {\bibfnamefont {P.}~\bibnamefont {Alonso-Gonz{\'a}lez}}, \bibinfo {author}
  {\bibfnamefont {K.}~\bibnamefont {Watanabe}}, \bibinfo {author}
  {\bibfnamefont {T.}~\bibnamefont {Taniguchi}}, \bibinfo {author}
  {\bibfnamefont {J.}~\bibnamefont {Hone}}, \bibinfo {author} {\bibfnamefont
  {R.}~\bibnamefont {Hillenbrand}},\ and\ \bibinfo {author} {\bibfnamefont
  {F.~H.}\ \bibnamefont {Koppens}},\ }\href {https://doi.org/10.1038/nmat4755}
  {\bibfield  {journal} {\bibinfo  {journal} {Nature Materials}\ }\textbf
  {\bibinfo {volume} {16}},\ \bibinfo {pages} {204} (\bibinfo {year}
  {2016})}\BibitemShut {NoStop}%
\bibitem [{\citenamefont {Woessner}\ \emph {et~al.}(2014)\citenamefont
  {Woessner}, \citenamefont {Lundeberg}, \citenamefont {Gao}, \citenamefont
  {Principi}, \citenamefont {Alonso-Gonz{\'a}lez}, \citenamefont {Carrega},
  \citenamefont {Watanabe}, \citenamefont {Taniguchi}, \citenamefont {Vignale},
  \citenamefont {Polini} \emph {et~al.}}]{Woessner2014}%
  \BibitemOpen
  \bibfield  {author} {\bibinfo {author} {\bibfnamefont {A.}~\bibnamefont
  {Woessner}}, \bibinfo {author} {\bibfnamefont {M.~B.}\ \bibnamefont
  {Lundeberg}}, \bibinfo {author} {\bibfnamefont {Y.}~\bibnamefont {Gao}},
  \bibinfo {author} {\bibfnamefont {A.}~\bibnamefont {Principi}}, \bibinfo
  {author} {\bibfnamefont {P.}~\bibnamefont {Alonso-Gonz{\'a}lez}}, \bibinfo
  {author} {\bibfnamefont {M.}~\bibnamefont {Carrega}}, \bibinfo {author}
  {\bibfnamefont {K.}~\bibnamefont {Watanabe}}, \bibinfo {author}
  {\bibfnamefont {T.}~\bibnamefont {Taniguchi}}, \bibinfo {author}
  {\bibfnamefont {G.}~\bibnamefont {Vignale}}, \bibinfo {author} {\bibfnamefont
  {M.}~\bibnamefont {Polini}}, \emph {et~al.},\ }\href
  {https://doi.org/10.1038/nmat4169} {\bibfield  {journal} {\bibinfo  {journal}
  {Nat. Mater.}\ }\textbf {\bibinfo {volume} {14}},\ \bibinfo {pages} {421}
  (\bibinfo {year} {2014})}\BibitemShut {NoStop}%
\bibitem [{\citenamefont {Fei}\ \emph {et~al.}(2012)\citenamefont {Fei},
  \citenamefont {Rodin}, \citenamefont {Andreev}, \citenamefont {Bao},
  \citenamefont {McLeod}, \citenamefont {Wagner}, \citenamefont {Zhang},
  \citenamefont {Zhao}, \citenamefont {Thiemens}, \citenamefont {Dominguez}
  \emph {et~al.}}]{Fei2012}%
  \BibitemOpen
  \bibfield  {author} {\bibinfo {author} {\bibfnamefont {Z.}~\bibnamefont
  {Fei}}, \bibinfo {author} {\bibfnamefont {A.}~\bibnamefont {Rodin}}, \bibinfo
  {author} {\bibfnamefont {G.~O.}\ \bibnamefont {Andreev}}, \bibinfo {author}
  {\bibfnamefont {W.}~\bibnamefont {Bao}}, \bibinfo {author} {\bibfnamefont
  {A.}~\bibnamefont {McLeod}}, \bibinfo {author} {\bibfnamefont
  {M.}~\bibnamefont {Wagner}}, \bibinfo {author} {\bibfnamefont
  {L.}~\bibnamefont {Zhang}}, \bibinfo {author} {\bibfnamefont
  {Z.}~\bibnamefont {Zhao}}, \bibinfo {author} {\bibfnamefont {M.}~\bibnamefont
  {Thiemens}}, \bibinfo {author} {\bibfnamefont {G.}~\bibnamefont {Dominguez}},
  \emph {et~al.},\ }\href {https://doi.org/10.1038/nature11253} {\bibfield
  {journal} {\bibinfo  {journal} {Nature}\ }\textbf {\bibinfo {volume} {487}},\
  \bibinfo {pages} {82} (\bibinfo {year} {2012})}\BibitemShut {NoStop}%
\bibitem [{\citenamefont {Fei}\ \emph {et~al.}(2015)\citenamefont {Fei},
  \citenamefont {Iwinski}, \citenamefont {Ni}, \citenamefont {Zhang},
  \citenamefont {Bao}, \citenamefont {Rodin}, \citenamefont {Lee},
  \citenamefont {Wagner}, \citenamefont {Liu}, \citenamefont {Dai} \emph
  {et~al.}}]{Fei2015}%
  \BibitemOpen
  \bibfield  {author} {\bibinfo {author} {\bibfnamefont {Z.}~\bibnamefont
  {Fei}}, \bibinfo {author} {\bibfnamefont {E.}~\bibnamefont {Iwinski}},
  \bibinfo {author} {\bibfnamefont {G.}~\bibnamefont {Ni}}, \bibinfo {author}
  {\bibfnamefont {L.}~\bibnamefont {Zhang}}, \bibinfo {author} {\bibfnamefont
  {W.}~\bibnamefont {Bao}}, \bibinfo {author} {\bibfnamefont {A.}~\bibnamefont
  {Rodin}}, \bibinfo {author} {\bibfnamefont {Y.}~\bibnamefont {Lee}}, \bibinfo
  {author} {\bibfnamefont {M.}~\bibnamefont {Wagner}}, \bibinfo {author}
  {\bibfnamefont {M.}~\bibnamefont {Liu}}, \bibinfo {author} {\bibfnamefont
  {S.}~\bibnamefont {Dai}}, \emph {et~al.},\ }\href
  {https://doi.org/10.1021/acs.nanolett.5b00912} {\bibfield  {journal}
  {\bibinfo  {journal} {Nano Letters}\ }\textbf {\bibinfo {volume} {15}},\
  \bibinfo {pages} {4973} (\bibinfo {year} {2015})}\BibitemShut {NoStop}%
\bibitem [{\citenamefont {Dai}\ \emph {et~al.}(2014)\citenamefont {Dai},
  \citenamefont {Fei}, \citenamefont {Ma}, \citenamefont {Rodin}, \citenamefont
  {Wagner}, \citenamefont {McLeod}, \citenamefont {Liu}, \citenamefont
  {Gannett}, \citenamefont {Regan}, \citenamefont {Watanabe} \emph
  {et~al.}}]{Dai2014}%
  \BibitemOpen
  \bibfield  {author} {\bibinfo {author} {\bibfnamefont {S.}~\bibnamefont
  {Dai}}, \bibinfo {author} {\bibfnamefont {Z.}~\bibnamefont {Fei}}, \bibinfo
  {author} {\bibfnamefont {Q.}~\bibnamefont {Ma}}, \bibinfo {author}
  {\bibfnamefont {A.}~\bibnamefont {Rodin}}, \bibinfo {author} {\bibfnamefont
  {M.}~\bibnamefont {Wagner}}, \bibinfo {author} {\bibfnamefont
  {A.}~\bibnamefont {McLeod}}, \bibinfo {author} {\bibfnamefont
  {M.}~\bibnamefont {Liu}}, \bibinfo {author} {\bibfnamefont {W.}~\bibnamefont
  {Gannett}}, \bibinfo {author} {\bibfnamefont {W.}~\bibnamefont {Regan}},
  \bibinfo {author} {\bibfnamefont {K.}~\bibnamefont {Watanabe}}, \emph
  {et~al.},\ }\href {https://doi.org/10.1126/science.1246833} {\bibfield
  {journal} {\bibinfo  {journal} {Science}\ }\textbf {\bibinfo {volume}
  {343}},\ \bibinfo {pages} {1125} (\bibinfo {year} {2014})}\BibitemShut
  {NoStop}%
\bibitem [{\citenamefont {Andersen}\ \emph {et~al.}(2015)\citenamefont
  {Andersen}, \citenamefont {Latini},\ and\ \citenamefont
  {Thygesen}}]{Andersen2015}%
  \BibitemOpen
  \bibfield  {author} {\bibinfo {author} {\bibfnamefont {K.}~\bibnamefont
  {Andersen}}, \bibinfo {author} {\bibfnamefont {S.}~\bibnamefont {Latini}},\
  and\ \bibinfo {author} {\bibfnamefont {K.~S.}\ \bibnamefont {Thygesen}},\
  }\href {https://doi.org/10.1021/acs.nanolett.5b01251} {\bibfield  {journal}
  {\bibinfo  {journal} {Nano Lett.}\ }\textbf {\bibinfo {volume} {15}},\
  \bibinfo {pages} {4616} (\bibinfo {year} {2015})}\BibitemShut {NoStop}%
\bibitem [{\citenamefont {Gjerding}\ \emph {et~al.}(2020)\citenamefont
  {Gjerding}, \citenamefont {Cavalcante}, \citenamefont {Chaves},\ and\
  \citenamefont {Thygesen}}]{Gjerding2020}%
  \BibitemOpen
  \bibfield  {author} {\bibinfo {author} {\bibfnamefont {M.~N.}\ \bibnamefont
  {Gjerding}}, \bibinfo {author} {\bibfnamefont {L.~S.~R.}\ \bibnamefont
  {Cavalcante}}, \bibinfo {author} {\bibfnamefont {A.}~\bibnamefont {Chaves}},\
  and\ \bibinfo {author} {\bibfnamefont {K.~S.}\ \bibnamefont {Thygesen}},\
  }\href {https://doi.org/10.1021/acs.jpcc.0c01635} {\bibfield  {journal}
  {\bibinfo  {journal} {J. Phys. Chem.}\ }\textbf {\bibinfo {volume} {124}},\
  \bibinfo {pages} {11609} (\bibinfo {year} {2020})}\BibitemShut {NoStop}%
\bibitem [{\citenamefont {Fetter}(2003)}]{Fetter2003}%
  \BibitemOpen
  \bibfield  {author} {\bibinfo {author} {\bibfnamefont {A.~L.}\ \bibnamefont
  {Fetter}},\ }\href
  {https://www.ebook.de/de/product/2192197/alexander_l_fetter_quantum_theory_of_many_particle_sys.html}
  {\emph {\bibinfo {title} {Quantum Theory of Many-Particle Systems}}}\
  (\bibinfo  {publisher} {Dover Publications Inc.},\ \bibinfo {year}
  {2003})\BibitemShut {NoStop}%
\bibitem [{\citenamefont {Hwang}\ and\ \citenamefont
  {Sarma}(2007)}]{Hwang2007}%
  \BibitemOpen
  \bibfield  {author} {\bibinfo {author} {\bibfnamefont {E.~H.}\ \bibnamefont
  {Hwang}}\ and\ \bibinfo {author} {\bibfnamefont {S.~D.}\ \bibnamefont
  {Sarma}},\ }\href {https://doi.org/10.1103/physrevb.75.205418} {\bibfield
  {journal} {\bibinfo  {journal} {Phys. Rev. B}\ }\textbf {\bibinfo {volume}
  {75}},\ \bibinfo {pages} {075418} (\bibinfo {year} {2007})}\BibitemShut
  {NoStop}%
\bibitem [{\citenamefont {Wunsch}\ \emph {et~al.}(2006)\citenamefont {Wunsch},
  \citenamefont {Stauber}, \citenamefont {Sols},\ and\ \citenamefont
  {Guinea}}]{Wunsch2006}%
  \BibitemOpen
  \bibfield  {author} {\bibinfo {author} {\bibfnamefont {B.}~\bibnamefont
  {Wunsch}}, \bibinfo {author} {\bibfnamefont {T.}~\bibnamefont {Stauber}},
  \bibinfo {author} {\bibfnamefont {F.}~\bibnamefont {Sols}},\ and\ \bibinfo
  {author} {\bibfnamefont {F.}~\bibnamefont {Guinea}},\ }\href
  {https://doi.org/10.1088/1367-2630/8/12/318} {\bibfield  {journal} {\bibinfo
  {journal} {New Journal of Physics}\ }\textbf {\bibinfo {volume} {8}},\
  \bibinfo {pages} {318} (\bibinfo {year} {2006})}\BibitemShut {NoStop}%
\bibitem [{\citenamefont {Principi}\ \emph {et~al.}(2009)\citenamefont
  {Principi}, \citenamefont {Polini},\ and\ \citenamefont
  {Vignale}}]{Principi2009}%
  \BibitemOpen
  \bibfield  {author} {\bibinfo {author} {\bibfnamefont {A.}~\bibnamefont
  {Principi}}, \bibinfo {author} {\bibfnamefont {M.}~\bibnamefont {Polini}},\
  and\ \bibinfo {author} {\bibfnamefont {G.}~\bibnamefont {Vignale}},\ }\href
  {https://doi.org/10.1103/physrevb.80.075418} {\bibfield  {journal} {\bibinfo
  {journal} {Physical Review B}\ }\textbf {\bibinfo {volume} {80}},\ \bibinfo
  {pages} {075418} (\bibinfo {year} {2009})}\BibitemShut {NoStop}%
\bibitem [{\citenamefont {Liu}\ and\ \citenamefont
  {Willis}(2010)}]{liu2010plasmon}%
  \BibitemOpen
  \bibfield  {author} {\bibinfo {author} {\bibfnamefont {Y.}~\bibnamefont
  {Liu}}\ and\ \bibinfo {author} {\bibfnamefont {R.~F.}\ \bibnamefont
  {Willis}},\ }\href
  {https://doi.org/https://doi.org/10.1103/PhysRevB.81.081406} {\bibfield
  {journal} {\bibinfo  {journal} {Physical Review B}\ }\textbf {\bibinfo
  {volume} {81}},\ \bibinfo {pages} {081406} (\bibinfo {year}
  {2010})}\BibitemShut {NoStop}%
\bibitem [{\citenamefont {Ong}\ and\ \citenamefont
  {Fischetti}(2012)}]{ong2012theory}%
  \BibitemOpen
  \bibfield  {author} {\bibinfo {author} {\bibfnamefont {Z.-Y.}\ \bibnamefont
  {Ong}}\ and\ \bibinfo {author} {\bibfnamefont {M.~V.}\ \bibnamefont
  {Fischetti}},\ }\href
  {https://doi.org/https://doi.org/10.1103/PhysRevB.86.165422} {\bibfield
  {journal} {\bibinfo  {journal} {Physical Review B}\ }\textbf {\bibinfo
  {volume} {86}},\ \bibinfo {pages} {165422} (\bibinfo {year}
  {2012})}\BibitemShut {NoStop}%
\bibitem [{\citenamefont {Fei}\ \emph {et~al.}(2011)\citenamefont {Fei},
  \citenamefont {Andreev}, \citenamefont {Bao}, \citenamefont {Zhang},
  \citenamefont {McLeod}, \citenamefont {Wang}, \citenamefont {Stewart},
  \citenamefont {Zhao}, \citenamefont {Dominguez}, \citenamefont {Thiemens}
  \emph {et~al.}}]{Fei2011}%
  \BibitemOpen
  \bibfield  {author} {\bibinfo {author} {\bibfnamefont {Z.}~\bibnamefont
  {Fei}}, \bibinfo {author} {\bibfnamefont {G.~O.}\ \bibnamefont {Andreev}},
  \bibinfo {author} {\bibfnamefont {W.}~\bibnamefont {Bao}}, \bibinfo {author}
  {\bibfnamefont {L.~M.}\ \bibnamefont {Zhang}}, \bibinfo {author}
  {\bibfnamefont {A.~S.}\ \bibnamefont {McLeod}}, \bibinfo {author}
  {\bibfnamefont {C.}~\bibnamefont {Wang}}, \bibinfo {author} {\bibfnamefont
  {M.~K.}\ \bibnamefont {Stewart}}, \bibinfo {author} {\bibfnamefont
  {Z.}~\bibnamefont {Zhao}}, \bibinfo {author} {\bibfnamefont {G.}~\bibnamefont
  {Dominguez}}, \bibinfo {author} {\bibfnamefont {M.}~\bibnamefont {Thiemens}},
  \emph {et~al.},\ }\href {https://doi.org/10.1021/nl202362d} {\bibfield
  {journal} {\bibinfo  {journal} {Nano Letters}\ }\textbf {\bibinfo {volume}
  {11}},\ \bibinfo {pages} {4701} (\bibinfo {year} {2011})}\BibitemShut
  {NoStop}%
\bibitem [{\citenamefont {Principi}\ \emph {et~al.}(2013)\citenamefont
  {Principi}, \citenamefont {Vignale}, \citenamefont {Carrega},\ and\
  \citenamefont {Polini}}]{Principi2013}%
  \BibitemOpen
  \bibfield  {author} {\bibinfo {author} {\bibfnamefont {A.}~\bibnamefont
  {Principi}}, \bibinfo {author} {\bibfnamefont {G.}~\bibnamefont {Vignale}},
  \bibinfo {author} {\bibfnamefont {M.}~\bibnamefont {Carrega}},\ and\ \bibinfo
  {author} {\bibfnamefont {M.}~\bibnamefont {Polini}},\ }\href
  {https://doi.org/10.1103/physrevb.88.121405} {\bibfield  {journal} {\bibinfo
  {journal} {Physical Review B}\ }\textbf {\bibinfo {volume} {88}},\ \bibinfo
  {pages} {121405} (\bibinfo {year} {2013})}\BibitemShut {NoStop}%
\bibitem [{\citenamefont {Langer}\ \emph {et~al.}(2010)\citenamefont {Langer},
  \citenamefont {Baringhaus}, \citenamefont {Pfnür}, \citenamefont
  {Schumacher},\ and\ \citenamefont {Tegenkamp}}]{Langer2010}%
  \BibitemOpen
  \bibfield  {author} {\bibinfo {author} {\bibfnamefont {T.}~\bibnamefont
  {Langer}}, \bibinfo {author} {\bibfnamefont {J.}~\bibnamefont {Baringhaus}},
  \bibinfo {author} {\bibfnamefont {H.}~\bibnamefont {Pfnür}}, \bibinfo
  {author} {\bibfnamefont {H.~W.}\ \bibnamefont {Schumacher}},\ and\ \bibinfo
  {author} {\bibfnamefont {C.}~\bibnamefont {Tegenkamp}},\ }\href
  {https://doi.org/10.1088/1367-2630/12/3/033017} {\bibfield  {journal}
  {\bibinfo  {journal} {New Journal of Physics}\ }\textbf {\bibinfo {volume}
  {12}},\ \bibinfo {pages} {033017} (\bibinfo {year} {2010})}\BibitemShut
  {NoStop}%
\bibitem [{\citenamefont {Cavalcante}\ \emph {et~al.}(2019)\citenamefont
  {Cavalcante}, \citenamefont {Gjerding}, \citenamefont {Chaves},\ and\
  \citenamefont {Thygesen}}]{cavalcante2019}%
  \BibitemOpen
  \bibfield  {author} {\bibinfo {author} {\bibfnamefont {L.~S.}\ \bibnamefont
  {Cavalcante}}, \bibinfo {author} {\bibfnamefont {M.~N.}\ \bibnamefont
  {Gjerding}}, \bibinfo {author} {\bibfnamefont {A.}~\bibnamefont {Chaves}},\
  and\ \bibinfo {author} {\bibfnamefont {K.~S.}\ \bibnamefont {Thygesen}},\
  }\href {https://doi.org/10.1021/acs.jpcc.9b04000} {\bibfield  {journal}
  {\bibinfo  {journal} {The Journal of Physical Chemistry C}\ }\textbf
  {\bibinfo {volume} {123}},\ \bibinfo {pages} {16373} (\bibinfo {year}
  {2019})}\BibitemShut {NoStop}%
\bibitem [{\citenamefont {Resta}(1994)}]{Resta1994}%
  \BibitemOpen
  \bibfield  {author} {\bibinfo {author} {\bibfnamefont {R.}~\bibnamefont
  {Resta}},\ }\href {https://doi.org/10.1103/revmodphys.66.899} {\bibfield
  {journal} {\bibinfo  {journal} {Reviews of Modern Physics}\ }\textbf
  {\bibinfo {volume} {66}},\ \bibinfo {pages} {899} (\bibinfo {year}
  {1994})}\BibitemShut {NoStop}%
\bibitem [{\citenamefont {King-Smith}\ and\ \citenamefont
  {Vanderbilt}(1993)}]{King-Smith1993}%
  \BibitemOpen
  \bibfield  {author} {\bibinfo {author} {\bibfnamefont {R.~D.}\ \bibnamefont
  {King-Smith}}\ and\ \bibinfo {author} {\bibfnamefont {D.}~\bibnamefont
  {Vanderbilt}},\ }\href {https://doi.org/10.1103/physrevb.47.1651} {\bibfield
  {journal} {\bibinfo  {journal} {Physical Review B}\ }\textbf {\bibinfo
  {volume} {47}},\ \bibinfo {pages} {1651} (\bibinfo {year}
  {1993})}\BibitemShut {NoStop}%
\bibitem [{Lin()}]{Link1}%
  \BibitemOpen
  \href@noop {} {\bibinfo {title} {The dielectric building blocks and
  $\text{QEH}$ software can be downloaded at
  https://cmr.fysik.dtu.dk/vdwh/vdwh.html}}\BibitemShut {NoStop}%
\bibitem [{\citenamefont {Yan}\ \emph {et~al.}(2013)\citenamefont {Yan},
  \citenamefont {Low}, \citenamefont {Zhu}, \citenamefont {Wu}, \citenamefont
  {Freitag}, \citenamefont {Li}, \citenamefont {Guinea}, \citenamefont
  {Avouris},\ and\ \citenamefont {Xia}}]{Yan2013}%
  \BibitemOpen
  \bibfield  {author} {\bibinfo {author} {\bibfnamefont {H.}~\bibnamefont
  {Yan}}, \bibinfo {author} {\bibfnamefont {T.}~\bibnamefont {Low}}, \bibinfo
  {author} {\bibfnamefont {W.}~\bibnamefont {Zhu}}, \bibinfo {author}
  {\bibfnamefont {Y.}~\bibnamefont {Wu}}, \bibinfo {author} {\bibfnamefont
  {M.}~\bibnamefont {Freitag}}, \bibinfo {author} {\bibfnamefont
  {X.}~\bibnamefont {Li}}, \bibinfo {author} {\bibfnamefont {F.}~\bibnamefont
  {Guinea}}, \bibinfo {author} {\bibfnamefont {P.}~\bibnamefont {Avouris}},\
  and\ \bibinfo {author} {\bibfnamefont {F.}~\bibnamefont {Xia}},\ }\href
  {https://doi.org/10.1038/nphoton.2013.57} {\bibfield  {journal} {\bibinfo
  {journal} {Nature Photonics}\ }\textbf {\bibinfo {volume} {7}},\ \bibinfo
  {pages} {394} (\bibinfo {year} {2013})}\BibitemShut {NoStop}%
\bibitem [{\citenamefont {McPherson}\ \emph {et~al.}(2003)\citenamefont
  {McPherson}, \citenamefont {Kim}, \citenamefont {Shanware},\ and\
  \citenamefont {Mogul}}]{McPherson2003}%
  \BibitemOpen
  \bibfield  {author} {\bibinfo {author} {\bibfnamefont {J.}~\bibnamefont
  {McPherson}}, \bibinfo {author} {\bibfnamefont {J.-Y.}\ \bibnamefont {Kim}},
  \bibinfo {author} {\bibfnamefont {A.}~\bibnamefont {Shanware}},\ and\
  \bibinfo {author} {\bibfnamefont {H.}~\bibnamefont {Mogul}},\ }\href
  {https://doi.org/10.1063/1.1565180} {\bibfield  {journal} {\bibinfo
  {journal} {Applied Physics Letters}\ }\textbf {\bibinfo {volume} {82}},\
  \bibinfo {pages} {2121} (\bibinfo {year} {2003})}\BibitemShut {NoStop}%
\bibitem [{\citenamefont {Cai}\ \emph {et~al.}(2007)\citenamefont {Cai},
  \citenamefont {Zhang}, \citenamefont {Zeng}, \citenamefont {Cheng},\ and\
  \citenamefont {Xu}}]{Cai2007}%
  \BibitemOpen
  \bibfield  {author} {\bibinfo {author} {\bibfnamefont {Y.}~\bibnamefont
  {Cai}}, \bibinfo {author} {\bibfnamefont {L.}~\bibnamefont {Zhang}}, \bibinfo
  {author} {\bibfnamefont {Q.}~\bibnamefont {Zeng}}, \bibinfo {author}
  {\bibfnamefont {L.}~\bibnamefont {Cheng}},\ and\ \bibinfo {author}
  {\bibfnamefont {Y.}~\bibnamefont {Xu}},\ }\href
  {https://doi.org/10.1016/j.ssc.2006.10.040} {\bibfield  {journal} {\bibinfo
  {journal} {Solid State Communications}\ }\textbf {\bibinfo {volume} {141}},\
  \bibinfo {pages} {262} (\bibinfo {year} {2007})}\BibitemShut {NoStop}%
\bibitem [{\citenamefont {Wagemann}\ \emph {et~al.}(2020)\citenamefont
  {Wagemann}, \citenamefont {Wang}, \citenamefont {Das},\ and\ \citenamefont
  {Mitra}}]{Wagemann2020}%
  \BibitemOpen
  \bibfield  {author} {\bibinfo {author} {\bibfnamefont {E.}~\bibnamefont
  {Wagemann}}, \bibinfo {author} {\bibfnamefont {Y.}~\bibnamefont {Wang}},
  \bibinfo {author} {\bibfnamefont {S.}~\bibnamefont {Das}},\ and\ \bibinfo
  {author} {\bibfnamefont {S.~K.}\ \bibnamefont {Mitra}},\ }\href
  {https://doi.org/10.1039/d0cp00200c} {\bibfield  {journal} {\bibinfo
  {journal} {Physical Chemistry Chemical Physics}\ }\textbf {\bibinfo {volume}
  {22}},\ \bibinfo {pages} {7710} (\bibinfo {year} {2020})}\BibitemShut
  {NoStop}%
\bibitem [{\citenamefont {Zhang}\ \emph {et~al.}(2015)\citenamefont {Zhang},
  \citenamefont {Qiao}, \citenamefont {Shi}, \citenamefont {Wu}, \citenamefont
  {Jiang},\ and\ \citenamefont {Tan}}]{Zhang2015a}%
  \BibitemOpen
  \bibfield  {author} {\bibinfo {author} {\bibfnamefont {X.}~\bibnamefont
  {Zhang}}, \bibinfo {author} {\bibfnamefont {X.-F.}\ \bibnamefont {Qiao}},
  \bibinfo {author} {\bibfnamefont {W.}~\bibnamefont {Shi}}, \bibinfo {author}
  {\bibfnamefont {J.-B.}\ \bibnamefont {Wu}}, \bibinfo {author} {\bibfnamefont
  {D.-S.}\ \bibnamefont {Jiang}},\ and\ \bibinfo {author} {\bibfnamefont
  {P.-H.}\ \bibnamefont {Tan}},\ }\href {https://doi.org/10.1039/c4cs00282b}
  {\bibfield  {journal} {\bibinfo  {journal} {Chemical Society Reviews}\
  }\textbf {\bibinfo {volume} {44}},\ \bibinfo {pages} {2757} (\bibinfo {year}
  {2015})}\BibitemShut {NoStop}%
\bibitem [{\citenamefont {Partoens}\ and\ \citenamefont
  {Peeters}(2006)}]{Partoens2006}%
  \BibitemOpen
  \bibfield  {author} {\bibinfo {author} {\bibfnamefont {B.}~\bibnamefont
  {Partoens}}\ and\ \bibinfo {author} {\bibfnamefont {F.~M.}\ \bibnamefont
  {Peeters}},\ }\href {https://doi.org/10.1103/physrevb.74.075404} {\bibfield
  {journal} {\bibinfo  {journal} {Physical Review B}\ }\textbf {\bibinfo
  {volume} {74}},\ \bibinfo {pages} {075404} (\bibinfo {year}
  {2006})}\BibitemShut {NoStop}%
\bibitem [{\citenamefont {Zhao}\ \emph {et~al.}(2013)\citenamefont {Zhao},
  \citenamefont {Ghorannevis}, \citenamefont {Amara}, \citenamefont {Pang},
  \citenamefont {Toh}, \citenamefont {Zhang}, \citenamefont {Kloc},
  \citenamefont {Tan},\ and\ \citenamefont {Eda}}]{Zhao2013}%
  \BibitemOpen
  \bibfield  {author} {\bibinfo {author} {\bibfnamefont {W.}~\bibnamefont
  {Zhao}}, \bibinfo {author} {\bibfnamefont {Z.}~\bibnamefont {Ghorannevis}},
  \bibinfo {author} {\bibfnamefont {K.~K.}\ \bibnamefont {Amara}}, \bibinfo
  {author} {\bibfnamefont {J.~R.}\ \bibnamefont {Pang}}, \bibinfo {author}
  {\bibfnamefont {M.}~\bibnamefont {Toh}}, \bibinfo {author} {\bibfnamefont
  {X.}~\bibnamefont {Zhang}}, \bibinfo {author} {\bibfnamefont
  {C.}~\bibnamefont {Kloc}}, \bibinfo {author} {\bibfnamefont {P.~H.}\
  \bibnamefont {Tan}},\ and\ \bibinfo {author} {\bibfnamefont {G.}~\bibnamefont
  {Eda}},\ }\href {https://doi.org/10.1039/c3nr03052k} {\bibfield  {journal}
  {\bibinfo  {journal} {Nanoscale}\ }\textbf {\bibinfo {volume} {5}},\ \bibinfo
  {pages} {9677} (\bibinfo {year} {2013})}\BibitemShut {NoStop}%
\bibitem [{\citenamefont {Molina-S{\'{a}}nchez}\ and\ \citenamefont
  {Wirtz}(2011)}]{MolinaSanchez2011}%
  \BibitemOpen
  \bibfield  {author} {\bibinfo {author} {\bibfnamefont {A.}~\bibnamefont
  {Molina-S{\'{a}}nchez}}\ and\ \bibinfo {author} {\bibfnamefont
  {L.}~\bibnamefont {Wirtz}},\ }\href
  {https://doi.org/10.1103/physrevb.84.155413} {\bibfield  {journal} {\bibinfo
  {journal} {Physical Review B}\ }\textbf {\bibinfo {volume} {84}},\ \bibinfo
  {pages} {155413} (\bibinfo {year} {2011})}\BibitemShut {NoStop}%
\bibitem [{\citenamefont {Peng}\ \emph {et~al.}(2016)\citenamefont {Peng},
  \citenamefont {Zhang}, \citenamefont {Shao}, \citenamefont {Xu},
  \citenamefont {Zhang},\ and\ \citenamefont {Zhu}}]{Peng2016}%
  \BibitemOpen
  \bibfield  {author} {\bibinfo {author} {\bibfnamefont {B.}~\bibnamefont
  {Peng}}, \bibinfo {author} {\bibfnamefont {H.}~\bibnamefont {Zhang}},
  \bibinfo {author} {\bibfnamefont {H.}~\bibnamefont {Shao}}, \bibinfo {author}
  {\bibfnamefont {Y.}~\bibnamefont {Xu}}, \bibinfo {author} {\bibfnamefont
  {X.}~\bibnamefont {Zhang}},\ and\ \bibinfo {author} {\bibfnamefont
  {H.}~\bibnamefont {Zhu}},\ }\href {https://doi.org/10.1039/c5ra19747c}
  {\bibfield  {journal} {\bibinfo  {journal} {{RSC} Advances}\ }\textbf
  {\bibinfo {volume} {6}},\ \bibinfo {pages} {5767} (\bibinfo {year}
  {2016})}\BibitemShut {NoStop}%
\bibitem [{\citenamefont {Berkdemir}\ \emph {et~al.}(2013)\citenamefont
  {Berkdemir}, \citenamefont {Guti{\'e}rrez}, \citenamefont
  {Botello-M{\'e}ndez}, \citenamefont {Perea-L{\'o}pez}, \citenamefont
  {El{\'\i}as}, \citenamefont {Chia}, \citenamefont {Wang}, \citenamefont
  {Crespi}, \citenamefont {L{\'o}pez-Ur{\'\i}as}, \citenamefont {Charlier}
  \emph {et~al.}}]{Berkdemir2013}%
  \BibitemOpen
  \bibfield  {author} {\bibinfo {author} {\bibfnamefont {A.}~\bibnamefont
  {Berkdemir}}, \bibinfo {author} {\bibfnamefont {H.~R.}\ \bibnamefont
  {Guti{\'e}rrez}}, \bibinfo {author} {\bibfnamefont {A.~R.}\ \bibnamefont
  {Botello-M{\'e}ndez}}, \bibinfo {author} {\bibfnamefont {N.}~\bibnamefont
  {Perea-L{\'o}pez}}, \bibinfo {author} {\bibfnamefont {A.~L.}\ \bibnamefont
  {El{\'\i}as}}, \bibinfo {author} {\bibfnamefont {C.-I.}\ \bibnamefont
  {Chia}}, \bibinfo {author} {\bibfnamefont {B.}~\bibnamefont {Wang}}, \bibinfo
  {author} {\bibfnamefont {V.~H.}\ \bibnamefont {Crespi}}, \bibinfo {author}
  {\bibfnamefont {F.}~\bibnamefont {L{\'o}pez-Ur{\'\i}as}}, \bibinfo {author}
  {\bibfnamefont {J.-C.}\ \bibnamefont {Charlier}}, \emph {et~al.},\ }\href
  {https://doi.org/10.1038/srep01755} {\bibfield  {journal} {\bibinfo
  {journal} {Scientific Reports}\ }\textbf {\bibinfo {volume} {3}},\ \bibinfo
  {pages} {1} (\bibinfo {year} {2013})}\BibitemShut {NoStop}%
\bibitem [{\citenamefont {Sengupta}\ \emph {et~al.}(2015)\citenamefont
  {Sengupta}, \citenamefont {Chanana},\ and\ \citenamefont
  {Mahapatra}}]{Sengupta2015}%
  \BibitemOpen
  \bibfield  {author} {\bibinfo {author} {\bibfnamefont {A.}~\bibnamefont
  {Sengupta}}, \bibinfo {author} {\bibfnamefont {A.}~\bibnamefont {Chanana}},\
  and\ \bibinfo {author} {\bibfnamefont {S.}~\bibnamefont {Mahapatra}},\ }\href
  {https://doi.org/10.1063/1.4907697} {\bibfield  {journal} {\bibinfo
  {journal} {{AIP} Advances}\ }\textbf {\bibinfo {volume} {5}},\ \bibinfo
  {pages} {027101} (\bibinfo {year} {2015})}\BibitemShut {NoStop}%
\end{thebibliography}%

\end{document}